\documentclass[11pt,a4paper]{article}

\usepackage{amsmath, amssymb}
\usepackage[utf8]{inputenc}
\usepackage{graphicx}
\usepackage[a4paper]{geometry} 
\usepackage{algorithm} 
\usepackage{algorithmicx}
\usepackage{algpseudocode} 
\usepackage{todonotes, multirow}
\usepackage[official]{eurosym}

\usepackage{etoolbox}

\makeatletter
\patchcmd{\@makecaption}
  {\parbox}
  {\advance\@tempdima-\fontdimen2} 
  {}{}
\makeatother

\usepackage[caption = false]{subfig} 
\usepackage{enumerate}
\usepackage{comment}
\usepackage{natbib, float}
\usepackage{todonotes}

\title{A Bayesian Nonparametric Analysis of the 2003 Outbreak of Highly Pathogenic Avian Influenza in the Netherlands}
\author{Rowland G. Seymour$^1$\footnote{To whom correspondence should be addressed. }$\>$, Theodore Kypraios$^1$, Philip D. O'Neill$^1$, Thomas J. Hagenaars$^2$}
\date{$^1$ School of Mathematical Sciences, University of Nottingham, UK \\ $^2$ Wageningen Bioveterinary Research (WBVR), Lelystad, The Netherlands}

\begin{document}
\maketitle

\begin{abstract}
Infectious diseases on farms pose both public and animal health risks, so understanding how they spread between farms is crucial for developing disease control strategies to prevent future outbreaks. We develop novel Bayesian nonparametric methodology to fit spatial stochastic transmission models in which the infection rate between any two farms is a function that depends on the distance between them, but without assuming a specified parametric form. Making nonparametric inference in this context is challenging since the likelihood function of the observed data is intractable because the underlying transmission process is unobserved. We adopt a fully Bayesian approach by assigning a transformed Gaussian Process prior distribution to the infection rate function, and then develop an efficient data augmentation Markov Chain Monte Carlo algorithm  to perform Bayesian inference. We use the posterior predictive distribution to simulate the effect of different disease control methods and their economic impact. We analyse a large outbreak of Avian Influenza in the Netherlands and infer the between-farm infection rate, as well as the unknown infection status of farms which were pre-emptively culled. We use our results to analyse ring-culling strategies, and conclude that although effective, ring-culling has limited impact in high density areas. 
\end{abstract}

\section{Introduction}\label{sec:intro}
Diseases of livestock and farmed poultry, such as Avian Influenza or Foot and Mouth Disease, pose serious public and animal health risks, as well as having a considerable impact on both the domestic and international farming economy. Authorities are keen to control the spread of such diseases as quickly as possible to reduce the health risks, but must also consider other stakeholders, such as farmers, and the economic consequences of intervention.

\begin{figure}
\centering
\includegraphics[width = 0.9\textwidth]{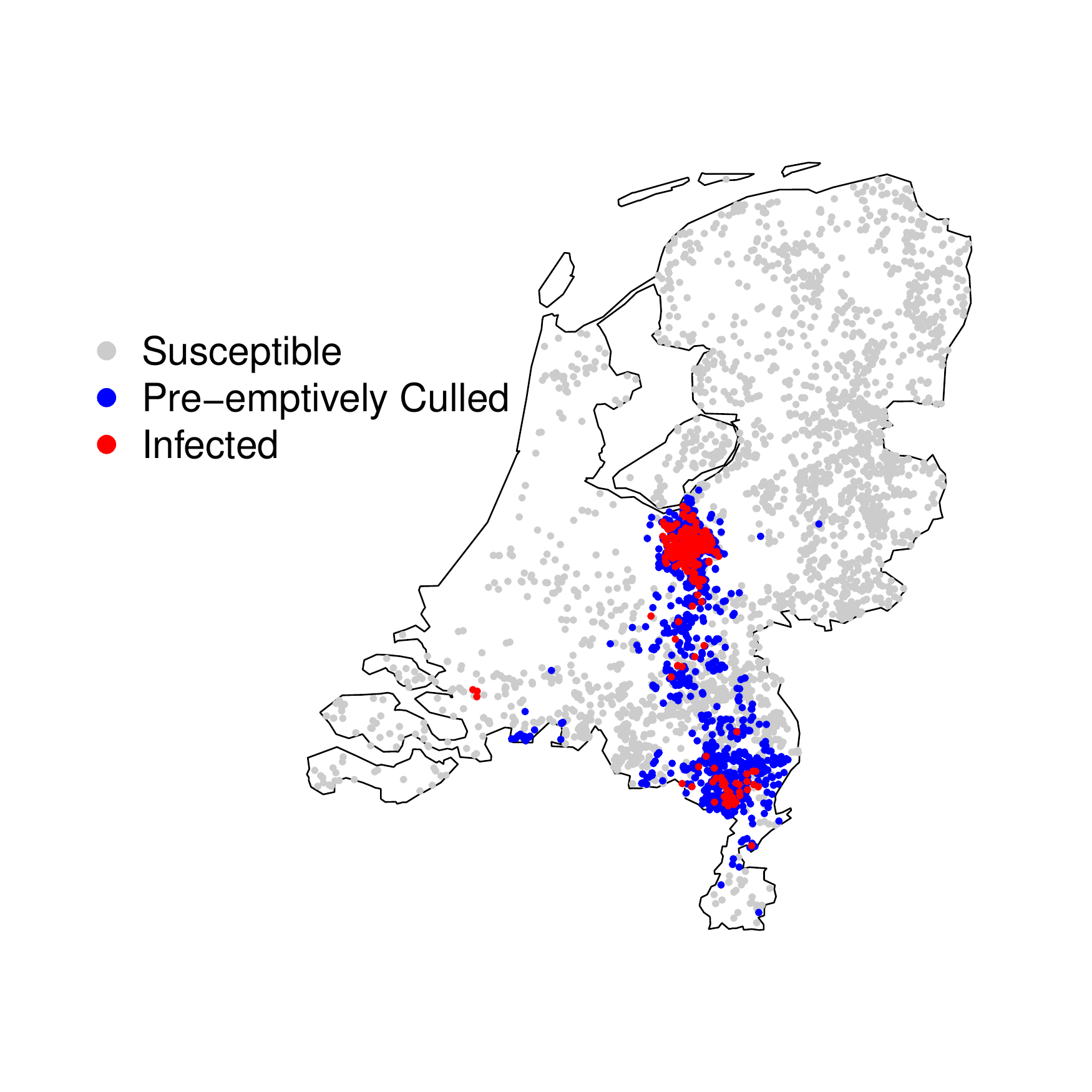}
\caption{\label{fig: map}A map of the poultry farms in the Netherlands with their status at the end of the outbreak.}
\end{figure}

In 2003 a serious outbreak of a highly pathogenic Avian Influenza A/H7N7 virus took place among poultry farms in the Netherlands. Over the course of three months, more than 30 millions birds were culled, 90 people developed influenza-like symptoms, with six confirmed cases, and one fatality occurred \citep{Koop04}. The Dutch authorities implemented a culling strategy to control the disease, whereby animals were culled on farms where the pathogen was detected, and pre-emptively culled on farms within a certain distance from the site of detection. For convenience we shall refer to farms as {\em naturally} culled or {\em pre-emptively} culled in the obvious manner. The culling strategy took place alongside strict biosecurity measures and a ban on the transportation of poultry goods \citep{DGSanco03}. In the data set we use there is a total of 5,397 Dutch poultry farms, including 241 infected farms and 1,232 pre-emptively culled farms. The approximate locations of the farms are shown in Figure \ref{fig: map}.

There is a clear spatial element to the spread of the disease; for example, there are two distinctive clusters of infected farms, within which there appears to be local transmission.  However, analysing the disease spread is challenging due to the fact that the times at which infections occurred are not observed. For farms which were confirmed to be infected, the date of poultry culling was recorded, but the date on which poultry on the farm were first infected was unobserved. The infection status of pre-emptively culled farms is considered uncertain, since absence of clinical suspicion at the time of culling would not necessarily rule out the presence of the pathogen.

Various data were collected during the outbreak. The particular data set that we shall focus on consists of the spatial coordinates of all poultry farms in the Netherlands, plus the culling dates and identities of all farms that were either naturally or pre-emptively culled. There are several previous approaches to modelling data from this outbreak. In \cite{Bov03}, the authors construct a model based on a generalised linear model proposed in \cite{Beck89b}, where the number of new infections per day is assumed to follow a Binomial distribution.  However, the infection rate is assumed to be constant between all farms, which is a questionable assumption given the clear spatial element to the spread of the disease. In \cite{Elb07}, the authors use a type of generalised linear model which allows for spatial variation in spread of the disease, and propose several plausible forms for the infection rate as a function of distance. It is assumed that pre-emptively culled farms are never infected, and that unobserved events such as infections occur at known times, obtained by simple assumptions motivated by expert opinion. Models are fitted using maximum likelihood methods and the Akaike Information Criterion is used to choose between them. In \cite{Back15}, the authors take a different approach modelling both within- and between-farm transmission. They model within-farm transmission using an SEIR (Susceptible-Exposed-Infective-Removed) model whose parameters are taken from the literature \citep[see, for example,][]{Goot05}. Transmission between farms is assumed to depend on the number of infectious animals and the distance between farms via a monotonically decreasing function in a similar fashion to the approach taken by \citet{Elb07}. The outbreak has also been studied from a public health and veterinary perspective, analysing the symptoms both humans and poultry display, see, for example, \cite{Fouch04, Koop04, Elb04}.

Although some previous modelling approaches attempt to capture the spatial variation in the infection rate, they rely on making strict parametric assumptions about the infection rate as a function of the distance between farms; such functions are commonly called distance kernels. The choice of a particular distance kernel may not accurately represent the underlying process and can lead to incorrect predictions which, in consequence, can have a significant impact on formulating policy decisions with regards to optimal disease control measures such as culling. Our approach removes the need to make such assumptions by modelling the infection rate nonparametrically. We do this by treating the infection rate as an unknown function with a transformed Gaussian Process (GP) prior distribution. This allows us to make more general assumptions about the type of function, for example how smooth it is, whether it is continuous, or if it is monotonic, rather than its exact shape. Furthermore, previous modelling approaches assume that the times at which farms were infected are known. In this paper we relax this assumption, by adopting a data-augmentation approach within a Bayesian framework in which we treat infection times as additional parameters. We make inference for the infection rate function using a Markov Chain Monte Carlo (MCMC) algorithm, which also allows us to infer the unobserved infection times, and to estimate the probability of any pre-emptively culled farm having been infected. We anticipate that the proposed framework is suitable for analysing completed major outbreaks among populations in which there is a clear spatial component in the infection rate.

The paper is structured as follows. In Section \ref{sec:methodology} we describe the available data, define our stochastic transmission model in detail and derive an augmented likelihood function assuming that the epidemic process is fully observed. In Section \ref{sec:inference} we present our Bayesian nonparametric approach by specifying a transformed  GP prior distribution for the infection rate function and the prior distributions for the other model parameters. We also describe an efficient MCMC algorithm to sample from the posterior distribution of the parameters given the observed data. In Section \ref{sec:results} we demonstrate the proposed models and methods via an application to simulated data and the Avian Influenza data set. We also illustrate how our methods can be used to assess control strategies. We finish in Section \ref{sec:conclusion} with brief conclusions and a discussion of our methods.

\section{Methodology} \label{sec:methodology}

\subsection{Data} \label{subsec:data}
The data set contains the geographical locations of 5,397 poultry farms in the Netherlands at the time of the outbreak. The data set lacks reliable information on small non-commercial flocks, as most of these are exempt from registration. For that reason, and because an earlier analysis showed that such ``backyard flocks'' played only a marginal role in this epidemic \citep{Bavinck2009}, we discarded all flocks with fewer than 500 animals from the dataset. For each farm, the data specifies its status at the end of the outbreak, describing whether or not it had contracted the virus, had been culled due to confirmed infection, or had been culled pre-emptively. For farms which were culled we have the date on which this occurred. After the removal of farms with fewer than 500 animals, the data set contains 4,466 farms. Of these, 233 farms were confirmed to be infected and consequently culled, while 1,232 farms were pre-emptively culled. Table \ref{tab:example data} illustrates the available information for each farm in the data set.

\begin{table}
\caption{\label{tab:example data} An example of the available information on each farm. Farms 1 and 2 were confirmed to be infected and culled in consequence. Farm 3 was culled pre-emptively and farm 4 was not culled. Farm geographical locations are provided in terms of $x$ and $y$ coordinates.}
\centering
	\begin{tabular}{|c|c|c|c|c|}\hline
	Farm ID & $x$ & $y$ & Culling Date &  Pre-emptively Culled\\ \hline
	1       & 5.25 & 52.13 & $5^{th}$ May  & $\times$ \\
	2       & 5.59 & 54.49 & $10^{th}$ April  & $\times$ \\
	3       & 4.99 & 55.00 & $2^{nd}$ May  & $\checkmark$ \\
	4       & 5.50 & 51.40 & -  & - \\
	$\vdots$      & $\vdots$ & $\vdots$ &  $\vdots$ & $\vdots$ \\
	\end{tabular}	
\end{table}

\subsection{Stochastic Epidemic Model}\label{sec:model}
We construct our model based on the standard SIR (Susceptible-Infective-Removed) epidemic model in continuous time; see, for example, \cite{Bail75, And00}. Consider a population consisting of $N$ farms. We assume that initially all farms are disease free apart from one which contains animals infected via some external source. At any time $t$, a farm is either {\em susceptible} to the disease, {\em infected} with the disease and infectious, or {\em removed} as the animals on the farm have been culled. The model dynamics can be separated into two processes: the infection process and the removal process. The infection process is governed by a rate function $\beta(d)$, where $d$ denotes the Euclidean distance between two farms. 

We assume an infectious farm infects a given susceptible farm that is $d$ km away according to a Poisson process with rate $\beta(d)$. The processes governing different pairs of farms are assumed to be independent. For the removal process, once a farm is infected it is assumed to be infectious for a time which follows a Gamma distribution, $\Gamma(\lambda, \gamma)$, which has mean $\lambda/\gamma$ and variance $\lambda/\gamma^2$. The infectious periods of different farms are assumed to be independent. Note that the infectious period of a farm is the time between infection and culling as a result of infection being detected, rather than the time period during which animals would be infectious in the absence of any intervention.

To account for the fact that some farms are pre-emptively culled by the authorities as a disease control measure,  we introduce pre-emptive culling times. We make no attempt to explicitly model the culling strategy, since in practice such strategies may change over time or not always be carried out as originally intended. Instead, we assume that pre-emptive cullings are deterministic events. 
If, under the disease control strategy, a farm is pre-emptively culled at time $t$, then the farm becomes removed at time $t$ irrespective of whether it is currently susceptible or infectious. From this time, it can no longer infect other farms or be infected. We shall refer to culling events that are not pre-emptive as natural cullings. The epidemic continues until there are no more infected farms. 

\subsection{Likelihood}
Recall that the observed data consist of culling times, which can be pre-emptive or not, and farm locations. To fit our model to such data in a Bayesian framework requires the likelihood of the observed data given the model parameters. However, such a likelihood is intractable in practice since its computation involves integrating over all unknown infection events; see, for example, \cite{On99, Jew09}. We therefore proceed by deriving a likelihood based on full observation of the epidemic process, and use a data-augmentation MCMC algorithm as described in Section 3. 

Let $N$ denote the total number of poultry farms in the Netherlands and $n$ the number of ever-infected farms.  We denote the infection and culling times for farm $j$ by $i_j$ and $r_j$ respectively, where culling may be pre-emptive or natural. We label the infected farms $1, \ldots, n$ by their culling date (i.e. $r_1 \leq r_2 \leq \ldots \leq r_n$) and the remaining farms $n + 1, \ldots, N$ arbitrarily. We denote by $\omega$ the label of the initially infected farm.

We denote by $\textbf{i} = \{i_1, \ldots, i_{\omega -1}, i_{\omega + 1}, \ldots, i_N\}$ the set of all infection times excluding the initial infection time $i_{\omega}$. If farm $j$ was not infected, its infection time is set to be $i_j = \infty$.  We account for pre-emptive culling by defining $r_j = \min(r^p_j, r^c_j)$, where $r_j^p$ and $r_j^c$ denote, respectively, the pre-emptive and natural culling time of farm $j$. We consider the times $r^p_j$ to be deterministic, and set $r^p_j = \infty$ if farm $j$ was not pre-emptively culled. For farms which were not culled at all, we set $r^p_j = r^c_j = \infty$, hence $r_j = \infty$. The sets  $\textbf{r}^c = \{r^c_1, \ldots, r^c_N\}$  and $\textbf{r}^p =\{r^p_1, \ldots, r^p_N\}$ denote the set of natural and pre-emptive culling times, respectively.

We require the following sets based on the infection status of the farms during the outbreak. Set $\mathcal{A}$ consists of the farms that remained susceptible to the disease throughout the course of the epidemic and were not culled, set $\mathcal{B}$ is the set of farms that were infected with the virus and naturally culled in consequence, set $\mathcal{C}$ is the set of farms that were infected but were culled pre-emptively, and finally set $\mathcal{D}$ consists of the farms that were not infected but still pre-emptively culled. These sets are shown in Table \ref{tab: farm sets}. Note that if a farm has been pre-emptively culled, we are unable to distinguish whether it belongs to set $\mathcal{C}$ or $\mathcal{D}$ unless its infection status is known.

\begin{table}
	\caption{\label{tab: farm sets} The infectious status of each farm at the end of the outbreak.}
\centering
	\begin{tabular}{|c|c|c|c|} \hline
	Set & Infected & Culled & Pre-emptively Culled \\ \hline
	$\mathcal{A}$ & $\times$ & $\times$ & $\times$ \\
	$\mathcal{B}$ & \checkmark & \checkmark & $\times$ \\
	$\mathcal{C}$ & \checkmark & \checkmark & \checkmark \\
	$\mathcal{D}$ & $\times$ & \checkmark & \checkmark  \\ \hline
	\end{tabular}
\end{table}

The likelihood function consists of three parts: a contribution from farms avoiding infection, a contribution from farms being infected, and a contribution from farms remaining infectious until culled. For a farm $k$ in either set $\mathcal{A}$, $\mathcal{B}$ or $\mathcal{C}$, the probability it avoids infection from infectious farm $j$, until either $j$ is removed or $k$ is infected, is 
$$
\psi_{j,k} = \exp\{-\beta(d_{j,k})((r_j \wedge i_k) -(i_j \wedge i_k))\},
$$
where $\beta(d)$ is the infection rate for a pair of farms that are $d$ km apart, and $a \wedge b = \min\{a, b\}$. The difference in minimum times is the amount of time during which farm $j$ is able to infect $k$. If farm $k$ is in set $\mathcal{D}$ we must take into account its pre-emptive culling time, $r_k = r^p_k$,  and the corresponding probability is given by 
$$
\psi_{j,k} = \exp\{-\beta(d_{j,k})((r_j \wedge r_k) -(i_j \wedge r_k))\}.
$$ 
When farm $j$ is infected, the set of farms that are able to infect $j$ is
$$
\mathcal{Y}_j = \{k: i_k < i_j < r_k \}, 
$$
so the event that $j$ is infected contributes to the likelihood function through the overall hazard rate of infection given by
$$
\phi_j = \sum\limits_{k \in \mathcal{Y}_j} \beta(d_{k,j}).
$$
For the removal process, the likelihood contribution is given by 
$$
\prod\limits_{j\in \mathcal{B}} p(r_j - i_j \mid \lambda, \gamma) \prod\limits_{j\in \mathcal{C}} S(r_j - i_j \mid \lambda, \gamma), 
$$
where $p(x \mid \lambda, \gamma)$ is the probability density function of a $\Gamma(\lambda, \gamma)$ distribution evaluated at $x$ and $S(x \mid \lambda, \gamma)$ is the survivor function 
$$
S(x \mid \lambda, \gamma) = \int\limits_{x}^\infty p(u \mid \lambda, \gamma) du.
$$
Farms in set $\mathcal{B}$, that were infected and culled at the end of their infectious period, contribute to the likelihood function through the total time during which they were infectious. For those in set $\mathcal{C}$, which were infected but culled pre-emptively, we consider their removal time as a censoring time, and compute the probability they would have remained infectious past their culling time. Combining the infection and removal processes gives the augmented likelihood function 
\begin{align}
&\pi(\textbf{i}, \textbf{r}^c, \mathcal{B}, \mathcal{C}, \mathcal{D}\mid\beta, \lambda, \gamma, \omega, i_\omega, \textbf{r}^p) \notag \\
&= \left(\prod\limits_{j \in \mathcal{B} \cup \mathcal{C}}\prod_{k=1}^N \psi_{j,k}\right)\left(\prod\limits_{\substack{j \in \mathcal{B} \cup \mathcal{C} \\ j \neq \omega}} \phi_j \right) \prod\limits_{j\in \mathcal{B}} p(r_j - i_j \mid \lambda, \gamma) \prod\limits_{j\in \mathcal{C}} S(r_j - i_j \mid \lambda, \gamma)  \label{eq: likelihood function} \\
&= \exp\left\{- \Psi \right\}\prod\limits_{\substack{j \in \mathcal{B} \cup \mathcal{C} \\ j \neq \omega}}\left( \sum\limits_{k \in \mathcal{Y}_j} \beta(d_{k, j})\right)\prod\limits_{j\in  \mathcal{B}} p(r_j - i_j \mid \lambda, \gamma) \prod\limits_{j\in \mathcal{C}} S(r_j - i_j \mid \lambda, \gamma),\notag 
\end{align}
where
\begin{align}
	\Psi = \sum\limits_{j \in \mathcal{B} \cup \mathcal{C}}\Bigg[&\sum\limits_{k\in \mathcal{A} \cup \mathcal{B} \cup \mathcal{C}} \beta(d_{j,k}) \left((r_j \wedge i_k) - (i_j \wedge i_k) \right)  \notag \\
	+ & \sum\limits_{k\in \mathcal{D}} \beta(d_{j,k}) \left((r_j \wedge r_k) - (i_j \wedge r_k) \right)\Bigg].
\label{eq: total infectious pressure}
\end{align}

Note that the set of culling times determines which farms belong to the set $\mathcal{A}$, which is why $\mathcal{A}$ does not appear explicitly in the left-hand side of equation (\ref{eq: likelihood function}).

\section{Bayesian Nonparametric Inference} \label{sec:inference}
We wish to make Bayesian inference for the unknown model parameters given the observed data of farm  locations and culling dates (see Table \ref{tab:example data}). If a farm was not  culled by the end of the outbreak, we assume that it remained susceptible throughout the outbreak. Hence, the observed culling dates determine which farms belong to set $\mathcal{A}$. For a farm that has been pre-emptively culled, its infection status is unknown and therefore we cannot determine from the observed data if such a farm  belongs to set $\mathcal{C}$ or $\mathcal{D}$. Also, the infection process defined in our model is not observed directly. Hence, the label of the initially infected farm $\omega$, its infection time $i_\omega$ and the infection times of the farms belonging to sets $\mathcal{B}$ or $\mathcal{C}$ are unknown. 

We adopt a data augmentation framework \citep[see, for example,][]{Jew09} in which we include the farms' unknown infection event times and statuses as additional model parameters to the ones which govern the transmission and removal processes. Combining the augmented data likelihood (\ref{eq: likelihood function}) with the joint prior distribution, by using Bayes' theorem, the target posterior density is given by

\begin{eqnarray*}
\pi(\beta, \gamma, \omega, i_\omega, \textbf{i}, \mathcal{C}, \mathcal{D}\mid \mathcal{B}, \lambda, \textbf{r}^p, \textbf{r}^c)  & \propto & \pi (\textbf{i}, \textbf{r}^c, \mathcal{B}, \mathcal{C}, \mathcal{D}\mid\beta, \lambda, \gamma, \omega, i_\omega, \textbf{r}^p) \nonumber \\  & & \times \pi(\beta) \pi(\lambda) \pi(i_\omega\mid\omega)\pi(\omega),
\end{eqnarray*}
where we have assumed that $\beta$, $\lambda$ and $(\omega, i_\omega)$ are independent \textit{a priori}.

\subsection{Prior Distributions}
We now discuss in detail the prior distributions for the infection rate function and the other model parameters.
\subsubsection{The Infection Rate Function}\label{sec:GP_prior}
We wish to infer the infection rate function $\beta$ nonparametrically and to do so we will use a transformed Gaussian Process (GP) as a prior distribution. We follow \cite{Ras06} and define a GP as a collection of points, any finite subset of which follow a multivariate Normal distribution. Suppose we wish to model a function $f$, over a space $\chi$, specifically being interested in the values of the function $f(x_1), \ldots, f(x_n)$ evaluated at the points $\textbf{x} = \{x_1, \ldots, x_n\}$. We specify the GP prior distribution on $f$ by 
$$
f \sim \mathcal{GP}(\mu, \Sigma), 
$$
where $\mu$ is the mean function and $\Sigma$ the covariance matrix, defined using a covariance function $k$. We build our assumptions about $f$ into the model through the covariance matrix, and to do so we use the squared exponential function. This is given by 
$$
\Sigma_{i,j} = k(x_i, x_j; \alpha, l), \quad k(x_i, x_j; \alpha, l) = \alpha^2\exp\left\{- \frac{(x_i - x_j)^2}{l^2}\right\}.
$$
The function $k$ has two hyperparameters, namely $\alpha$, which controls the overall variance, and $l$, which controls the length scale. The value of $l$ essentially determines how much the function can change as the input changes.  We implicitly assume that $f$ is smooth and differentiable.  Many other choices for the kernel function are available \citep{Ras06}, but our choice appears suitable for the application at hand.

The input space of the function $\beta$ is the space of Euclidean distances. We specifically wish to evaluate $\beta$ at $\boldsymbol{d}$, the set of pair-wise distances between all farms. As the GP prior distribution gives non-zero probability to negative values and we are modelling a rate which is always positive, we introduce a dummy function $g$ and use the transformation $\beta = \exp\{g\}$. In other words, we are placing a GP prior distribution on $\log \beta$ by specifying that
$$
g\sim \mathcal{GP}(0, \Sigma), \quad \Sigma_{ij} = k(d_{i}, d_{j};\, \alpha, l), \quad  \beta = \exp\{g\} 
$$
where $d_i$ is the Euclidean distance between the $i$th pair of farms.

A well-known problem arises with GPs when the size of the covariance matrix is large, since this creates computational difficulties with matrix inversion and decomposition; see, for example, \cite{Hens13, Csat02, Quin05}. For the Avian Influenza data set, there are over 9 million unique pair-wise distances from which the covariance matrix is constructed. In the MCMC algorithm we will develop, we will require the covariance matrix to be repeatedly decomposed and inverted, which is not feasible in practice with such a large matrix. We therefore approximate the GP prior distribution using a projection method first described in \cite{Quin05}. We construct a pseudo set of distances, $\bar{\textbf{d}}$, that is much smaller than the original set $\textbf{d}$. While $\bar{\textbf{d}}$ need not be a subset of $\textbf{d}$ it should provide an adequate representation of $\textbf{d}$.
We then place a GP prior distribution on the pseudo set and draw samples from this distribution. The joint prior distribution of the pseudo function, $\bar{f}$ and the full function $f$ is 
$$
\begin{pmatrix}
	\bar{f} \\ f	
\end{pmatrix} \sim 
\mathcal{GP}\left( \textbf{0}, \begin{pmatrix}
	\Sigma_{\bar{\textbf{d}}, \bar{\textbf{d}}} & \Sigma_{{\textbf{d}}, \bar{\textbf{d}}} \\
	\Sigma_{\bar{\textbf{d}}, {\textbf{d}}} & \Sigma_{\textbf{d}, \textbf{d}} 
\end{pmatrix}\right),
$$
where the subscripts on the $\Sigma$ matrices denote the vectors used to construct them. We can then project the samples onto the full data set by considering the conditional distribution of $f$ given the pseudo function $\bar{f}$. We take $f$ to be the mean of this distribution, which is given by
$$
f = \Sigma_{\textbf{d}, \bar{\textbf{d}}}\Sigma_{\bar{\textbf{d}}}^{-1}\bar{f},
$$
where $\Sigma_{\bar{\textbf{d}}}$ is the covariance matrix of the approximating prior distribution. Some care is needed to construct the pseudo set $\bar{\textbf{d}}$, because we must ensure the points are sufficient in number and suitably placed across the entire domain to capture the features of $\beta$. Simulation studies in \citet{Seymour20} suggest that the error introduced by this approximation is small, even when the number of pseudo distances is as small as 10\% of the total number of pairwise distances. This method assumes that the prior distribution over the pseudo data set has the same properties as that over the original data set, which is a reasonable assumption as they are both sets of Euclidean distances. 

The value of the length scale parameter $l$ can have a material impact on the results, and it is not obvious how to assign a suitable value. We therefore place a non-informative prior distribution on this parameter,specifically $l \sim \hbox{Exp}(0.01)$, where $\hbox{Exp}(a)$ denotes an exponential distribution with mean $a^{-1}$. Inferring both hyperparameters of the GP ($l$ and $\alpha$) can be very challenging \citep{Zhang2004}, and therefore we assume the variance parameter $\alpha$ is known \textit{a priori}.  We choose a value such that samples from the prior distribution are over a large enough range to capture the scale of the infection rate. 

\subsubsection{Other Model Parameters}
Recall that the infectious period distribution is assumed to be a $\Gamma(\lambda, \gamma)$ distribution. We follow \cite{Jew09} and assume that $\lambda$ is known and place an uninformative prior distribution on $\gamma$, specifically $\gamma \sim \hbox{Exp}(0.01)$.  

For the infection times, we place a discrete uniform prior distribution on the label of the initially infected farm $\omega$. We set a time axis by assuming the first culling to be at time zero, so that $r_1=0$, and set the prior distribution on the infection time of $\omega$ by 
$$
(i_\omega \mid \omega) = -z, \quad z \sim \hbox{Exp}(0.01). 
$$  
\subsection{Markov Chain Monte Carlo Algorithm} \label{sec:mcmc}
The density of the full posterior distribution is given by
\begin{align}
	&\pi(\beta, \gamma, \omega, i_\omega, \textbf{i}, \mathcal{C}, \mathcal{D}\mid \mathcal{B}, \lambda, \textbf{r}^{c}, \textbf{r}^p) \notag \\
	&\propto \exp\left\{- \Psi \right\}\prod\limits_{\substack{j=1 \\ j \neq \omega}}^n\left( \sum\limits_{k \in \mathcal{Y}_j} \exp\{g(d_{k, j})\}\right) \prod\limits_{j\in \mathcal{B}} p(r_j - i_j \mid \lambda, \gamma) \prod\limits_{j\in \mathcal{C}} S(r_j - i_j \mid \lambda, \gamma) \nonumber \\
	&\mbox{}\hspace{0.5cm}\times \mathcal{GP}(g; \, \textbf{0}, \Sigma)\exp\{-0.01l\} \exp\{-0.01\gamma\}  \exp\{0.01i_\omega\}. \notag
\end{align}
The likelihood function is the same as in equation (\ref{eq: likelihood function}) and $\Psi$ is given in equation (\ref{eq: total infectious pressure}) with $\beta$ replaced by the inferred value $\exp\{g\}$. The term $\mathcal{GP}(g; \, \textbf{0}, \Sigma)$ refers to the finite dimension form of the GP, which is the probability density function of a multivariate Gaussian distribution evaluated at $g(\textbf{d})$, with the corresponding mean vector $\boldsymbol{0}$ and covariance matrix $\Sigma$.  We cannot sample from the posterior distribution directly so construct an MCMC algorithm, which is shown in Algorithm \ref{alg: MCMC outline}. There are five main steps to the algorithm and these are described in detail below.

\begin{algorithm}[H]
\caption{Structure of the MCMC algorithm}
	\begin{algorithmic}[1]
		\State Initialise the chain with values $g^{(0)}$, $\gamma^{(0)}$, $l^{(0)}$, and $\textbf{i}^{(0)}$;
		\Statex \textit{Repeat the following steps:}
		\State Update $g$ using a Metropolis-Hastings step;
		\State Update $l$ using a random walk Metropolis-Hastings step;
		\State Update $\gamma$ using a random walk Metropolis-Hastings step;
		\State Update $\omega$ using a random walk Metropolis-Hastings step;
		\State Update $i_\omega$ using a random walk Metropolis-Hastings step;
		\State Choose one of the following steps with equal probability: 
		\begin{itemize}
			\item Update an infection time
			\item Remove an infection time for a pre-emptively culled farm
			\item Add an infection time for a pre-emptively culled farm
		\end{itemize}
	\end{algorithmic}
	\label{alg: MCMC outline}
\end{algorithm} 

\subsubsection{Updating the Infection Rate}
The first step is concerned with sampling the dummy function $g$, which we do using an underrelaxed proposal mechanism described in \cite{Neal98}. This allows us to update the function as a block while reducing computational complexity. Given the current function $g$, we propose a new function $g^\prime$ by 
$$
g^\prime(\textbf{d}) = \sqrt{1-\delta^2}g(\textbf{d}) + \delta\nu(\textbf{d}), \quad \nu \sim \mathcal{GP}(\boldsymbol{0}, \Sigma),
$$
where $\delta \in (0, 1]$ is a tuning parameter, $g(\boldsymbol{d})$ is the value of the function $g$ at the current iteration, and $\nu(\boldsymbol{d})$ is a sample drawn from the prior distribution $\mathcal{GP}(\boldsymbol{0}, \Sigma)$ where $\Sigma$ denotes the covariance matrix of the GP.  The computational advantage of this is that the prior ratio is the inverse of the proposal ratio so the Metropolis-Hastings acceptance probability reduces to the likelihood ratio (see Section 1 of the supplementary material)
$$
p_{acc} = \frac{\pi(\textbf{i}, \textbf{r}^c, \mathcal{B}, \mathcal{C}, \mathcal{D} \mid g^\prime, \lambda, \gamma, \omega, i_\omega, \textbf{r}^p)}{\pi(\textbf{i}, \textbf{r}^c, \mathcal{B}, \mathcal{C}, \mathcal{D}\mid g, \lambda, \gamma, \omega, i_\omega, \textbf{r}^p)} \wedge 1.
$$ 
When the projection approximation method is used, we first propose new values for $\bar{g}$ on the input space $\bar{\textbf{d}}$ using the above proposal, i.e. $\bar{g}'(\bar{\textbf{d}}) = \sqrt{1-\delta^2} \bar{g}(\bar{\textbf{d}}) + \delta \nu(\bar{\textbf{d})}$, and then project $\bar{g}'$ onto $g'$ which is then used in the Metropolis-Hastings ratio above.

\subsubsection{Updating $l$}
To update the GP prior distribution length scale, $l$, we use a Gaussian random walk Metropolis algorithm by first proposing a new length scale, $l'$, from $N(l, \sigma^2_l)$ where $\gamma$ is the current value and $\sigma_l^2$ is a tuning parameter, and then accept $l'$ with probability
$$
p_{acc} = \frac{\mathcal{GP}(g; \mathbf{0}, \Sigma_{l'})}{\mathcal{GP}(g; \mathbf{0}, \Sigma_{l})}\frac{\pi(l')}{\pi(l)} \wedge 1.
$$

\subsubsection{Updating $\gamma$}
To update $\gamma$, we also use a Gaussian random walk Metropolis algorithm by proposing $\gamma'$ from the distribution $N(\gamma, \sigma^2_\gamma)$ and accepting this with probability
$$
p_{acc} = \frac{\prod\limits_{j\in \mathcal{B}} p(r_j - i_j \mid \lambda, \gamma') \prod\limits_{j\in \mathcal{C}} S(r_j - i_j \mid \lambda, \gamma') }{\prod\limits_{j\in \mathcal{B}} p(r_j - i_j \mid \lambda, \gamma) \prod\limits_{j\in \mathcal{C}} S(r_j - i_j \mid \lambda, \gamma)}\frac{\pi(\gamma')}{\pi(\gamma)} \wedge 1.
$$

\subsubsection{Updating Infection Times}
The final step in the algorithm concerns the unobserved infection times. We use a method proposed in \cite{On99} and then further developed in \cite{Jew09}. We choose one of three actions with equal probability: (i) propose to move an existing infection time;  (ii) propose to add a new infection time; (iii) propose to delete a previously-added infection time.
 

(i) {\em \bf Updating an infection time} of a farm in sets $\mathcal{B}$ or $\mathcal{C}$ is the simplest of the three procedures. To do this, we randomly choose a farm $j$ that is currently infected and propose a new infection time by $i'_j = r_j - t_j$, where  $t_j \sim \Gamma(\lambda, \gamma)$ and $\gamma$ denotes the current value of the parameter in the chain. We accept $i'_j$ with probability
$$
p_{acc} = \frac{p(r_j - i_j\mid \lambda, \gamma)}{p(r_j - i'_j\mid \lambda, \gamma)}\frac{\pi(\textbf{i} - {i_j} + {i_j'}, \textbf{r}^c \mid g, \lambda, \gamma, i_\omega, \textbf{r}^p, \mathcal{B}, \mathcal{C}, \mathcal{D})}{\pi(\textbf{i}, \textbf{r}^c\mid g, \lambda, \gamma, i_\omega, \textbf{r}^p, \mathcal{B}, \mathcal{C}, \mathcal{D})} \wedge 1, 
$$
where $\textbf{i} - i_j + i'_j$ is the set \textbf{i} with $i_j$ removed and $i'_j$ included. 

(ii) When {\em \bf adding an infection time}, first define $m$ to be the number of pre-emptively culled farms. We suppose that at the current iteration of the algorithm, $\tilde{m}$ of the farms which were pre-emptively culled have had infection times added by the algorithm; that is farms belonging in set $\mathcal{C}$. We randomly choose one of the $m - \tilde{m}$ pre-emptively culled farms with no infection time and propose an infection time for it. If $m = \tilde{m}$, we abandon this step. We generate an infection time as above and accept it with probability
\begin{align*}
p_{acc} &= \frac{1/(\tilde{m}+1)}{(1/(m - \tilde{m}))p(r_j - i'_j\mid \lambda, \gamma)}\frac{\pi(\textbf{i} + {i_j} , \textbf{r}^c, \mathcal{C}, \mathcal{D} \mid g, \lambda, \gamma, i_\omega, \textbf{r}^p)}{\pi(\textbf{i}, \textbf{r}^c, \mathcal{C}, \mathcal{D} \mid g, \lambda, \gamma, i_\omega, \textbf{r}^p)} \wedge 1 \\
	&= \frac{m - \tilde{m}}{(\tilde{m}+1)p(r_j - i'_j\mid \lambda, \gamma)}\frac{\pi(\textbf{i} + {i_j} , \textbf{r}^c, \mathcal{C}, \mathcal{D} \mid g, \lambda, \gamma, i_\omega, \textbf{r}^p)}{\pi(\textbf{i}, \textbf{r}^c, \mathcal{C}, \mathcal{D} \mid g, \lambda, \gamma, i_\omega, \textbf{r}^p)} \wedge 1. \\
\end{align*}

(iii) Finally, if we choose to {\em \bf delete an infection time} for a pre-emptively culled farm, we randomly choose a pre-emptively culled farm $j$ which at the current iteration has an infection time added and we propose to remove its infection time. Should there be no farms with an unknown infection status, which, at the current iteration of the algorithm, have had an infection time added, the step is abandoned. We accept this proposal with probability 
\begin{align*}
	p_{acc} & = \frac{1/(m - (\tilde{m} - 1))p(r_j - i_j\mid \lambda, \gamma)}{1/\tilde{m}}\frac{\pi(\textbf{i} - {i_j} , \textbf{r}^c\mid g, \lambda, \gamma, i_\omega, \textbf{r}^p, \mathcal{B}, \mathcal{C}, \mathcal{D})}{\pi(\textbf{i}, \textbf{r}^c\mid g, \lambda, \gamma, i_\omega, \textbf{r}^p, \mathcal{B}, \mathcal{C}, \mathcal{D})}\wedge 1 \\
	&= \frac{p(r_j - i_j\mid \lambda, \gamma)\tilde{m}}{m - (\tilde{m} - 1)}\frac{\pi(\textbf{i} - {i_j} , \textbf{r}^c, \mathcal{B}, \mathcal{C}, \mathcal{D}\mid g, \lambda, \gamma, i_\omega, \textbf{r}^p)}{\pi(\textbf{i}, \textbf{r}^c, \mathcal{B}, \mathcal{C}, \mathcal{D}\mid g, \lambda, \gamma, i_\omega, \textbf{r}^p)}\wedge 1.
\end{align*}

\section{Results} \label{sec:results}
We now present the results of our method applied to two data sets. The first is a simulated data set and the second is the Avian Influenza data set described in Section \ref{sec:intro}. We then use the posterior predictive distribution to analyse the impact of various culling strategies for the Avian Influenza outbreak.

\subsection{Simulation Study} \label{sec:sim_study}
We generated the position of 1,000 farms uniformly at random on a square with side length 30km. We then simulated 250 outbreaks of Avian Influenza using the infection rate function
$$
\beta(d) = 0.6\exp\{-2d\}.
$$ 
The infectious period distribution parameters were $\lambda = 4$ and $\gamma = 0.8$. This gives a mean infectious period, which represents the time from infection to culling, of $\lambda / \gamma = 5$ days, suitable for influenza-like diseases among livestock. The simulations also included a deterministic culling strategy such that once a farm was culled following a positive test, all farms within a 1km radius were pre-emptively culled.  Note that although this strategy is inspired by what happened in the actual outbreak, it is somewhat idealised since, as mentioned in Section \ref{sec:model}, culling strategies may change over time.

We discarded simulated data sets with less than 100 infected farms, since our focus is towards analysing sizeable outbreaks, and any nonparametric modelling approach will struggle with a small data set. This left 175 data sets. Mimicking the data available in the Avian Influenza outbreak, for each simulated data set we assume that in addition to the coordinates we only observe the culling times and whether a farm has been pre-emptively culled or not.  The infectious period shape parameter is assumed to be $\lambda=4$. 

Estimating both the length scale ($l$) and the variance parameter ($\alpha$) can be very challenging \citep{Zhang2004} and computationally expensive \citep{chalupka2013}. It is therefore common to treat either parameter, or both of them, fixed and known. Care is indeed needed when specifying a value for the variance parameter $\alpha$. A very small value yields slow convergence times for the Markov chain. On the other hand, a very large value will lead to poor MCMC mixing. In practice, $\alpha$ needs to chosen such that the prior distribution for $\beta$ covers a large space, particularly $\beta(0)$, but not so large that the Markov chain mixes poorly. Therefore, we fix the length scale parameter as $l=3$ and the variance parameter $\alpha=9$ due to the computation time required to perform inference for both hyperparameters and infection times for pre-emptively culled farms for each of the simulated data sets. We also repeated the analysis for $l=2$ and $l=5$, and found that the results described below were essentially unchanged (see Section 2 of the supplementary material).

We fitted the model described in Section \ref{sec:methodology} by assuming that the infection rate is a function that depends only on the distance between farms and assigned a Gaussian Process prior distribution as described in Section \ref{sec:GP_prior}. Due to the number of farms, we used the GP approximation method, constructing the pseudo set of distances by taking 256 equally-spaced points from zero to the largest distance in the data set. We  employed the MCMC algorithm described in Section \ref{sec:mcmc} to fit the model to each of the 175 data sets.

\begin{figure}
    \centering
      \caption{Top left: The results of the nonparametric method for the infection rate in the simulated data sets. We report the median and 95\% credible interval of all 175 posterior medians. Top right: The distribution of the posterior median estimates for the mean infectious period. Bottom left: The distribution of the relative error in the sum of the infection times. Bottom right: The 175 posterior medians. \label{fig: simulation study posterior rates}}
  \includegraphics[width=.45\linewidth]{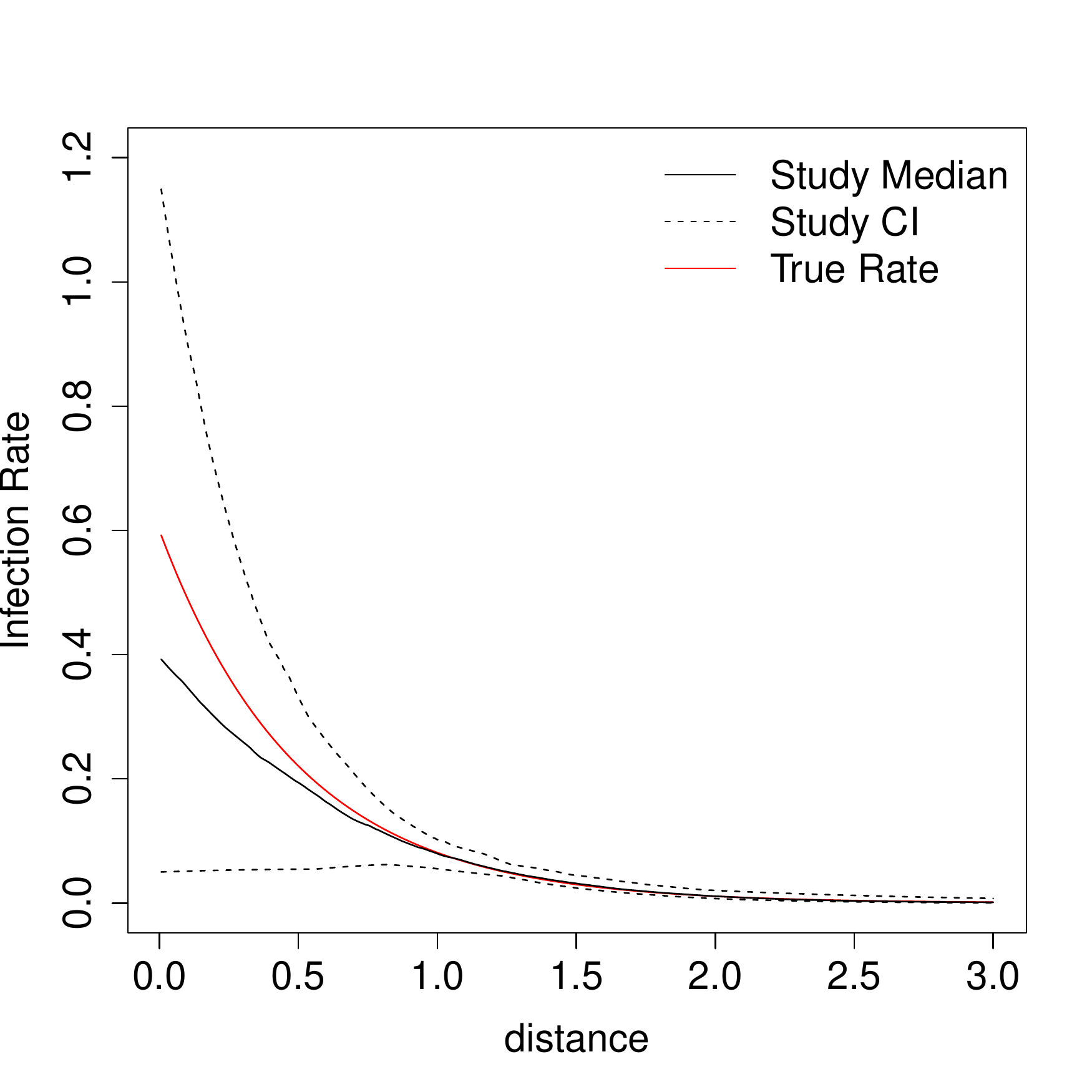}  
  \includegraphics[width=.45\linewidth]{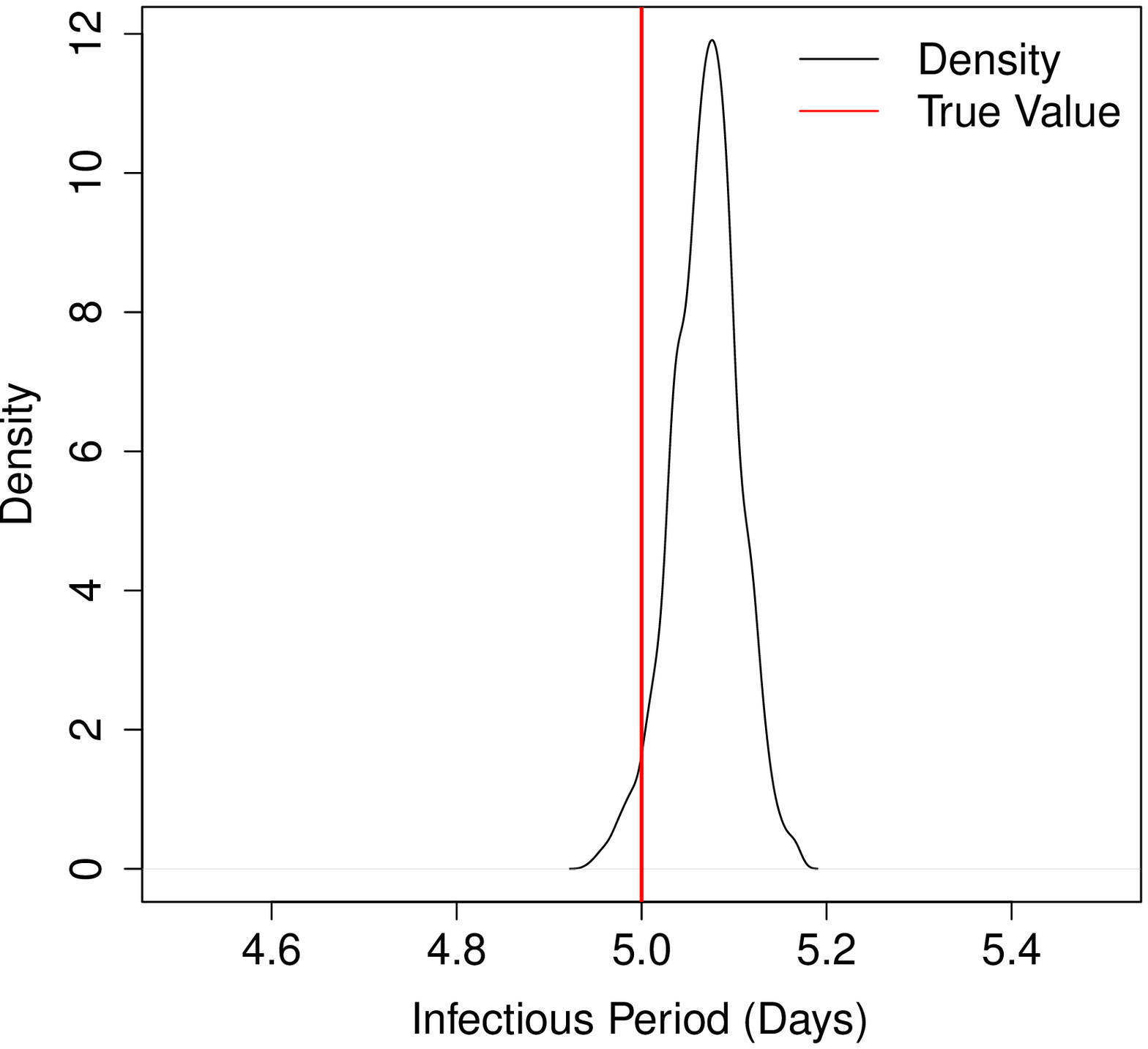}  
  \includegraphics[width=.45\linewidth]{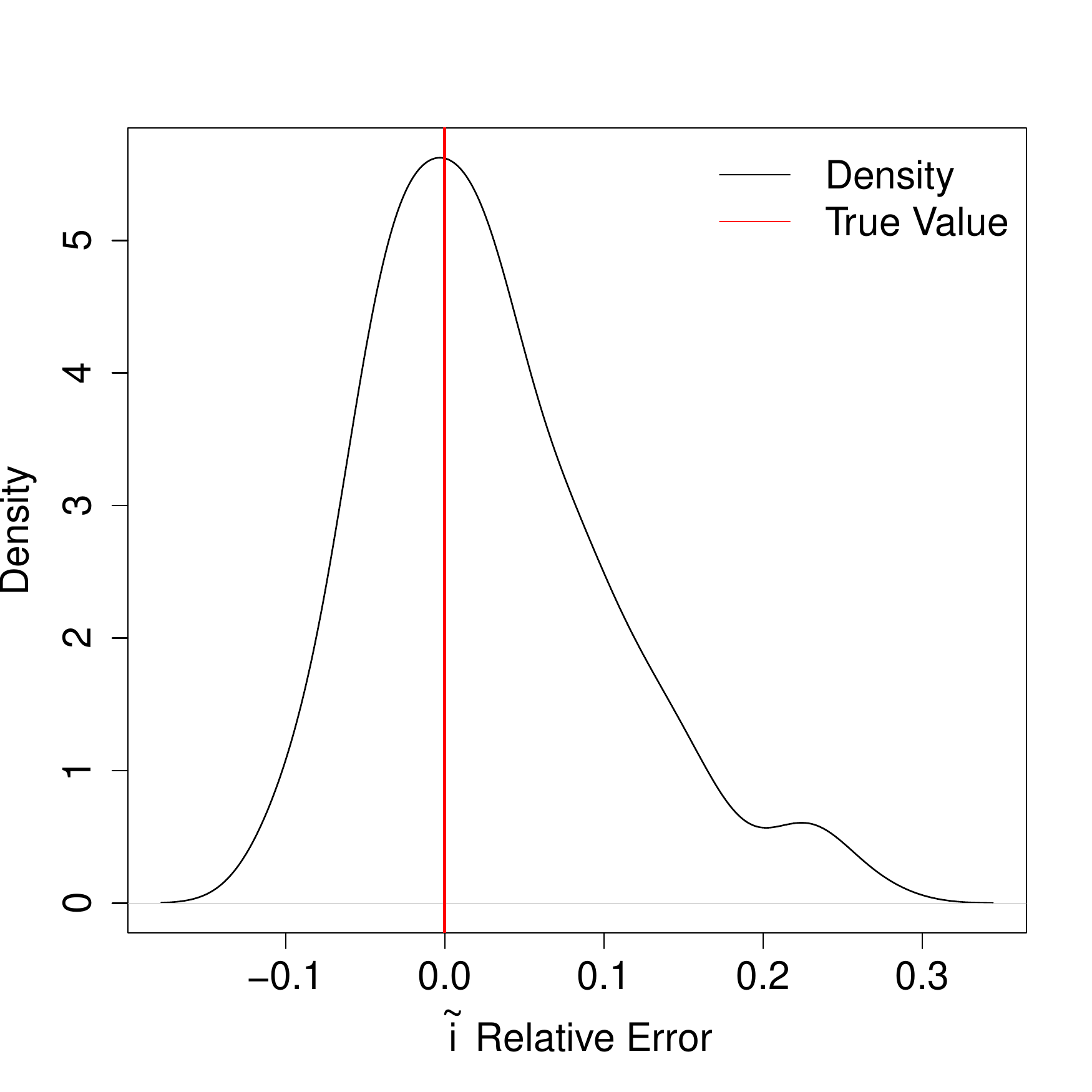}
  \includegraphics[width = 0.45\linewidth]{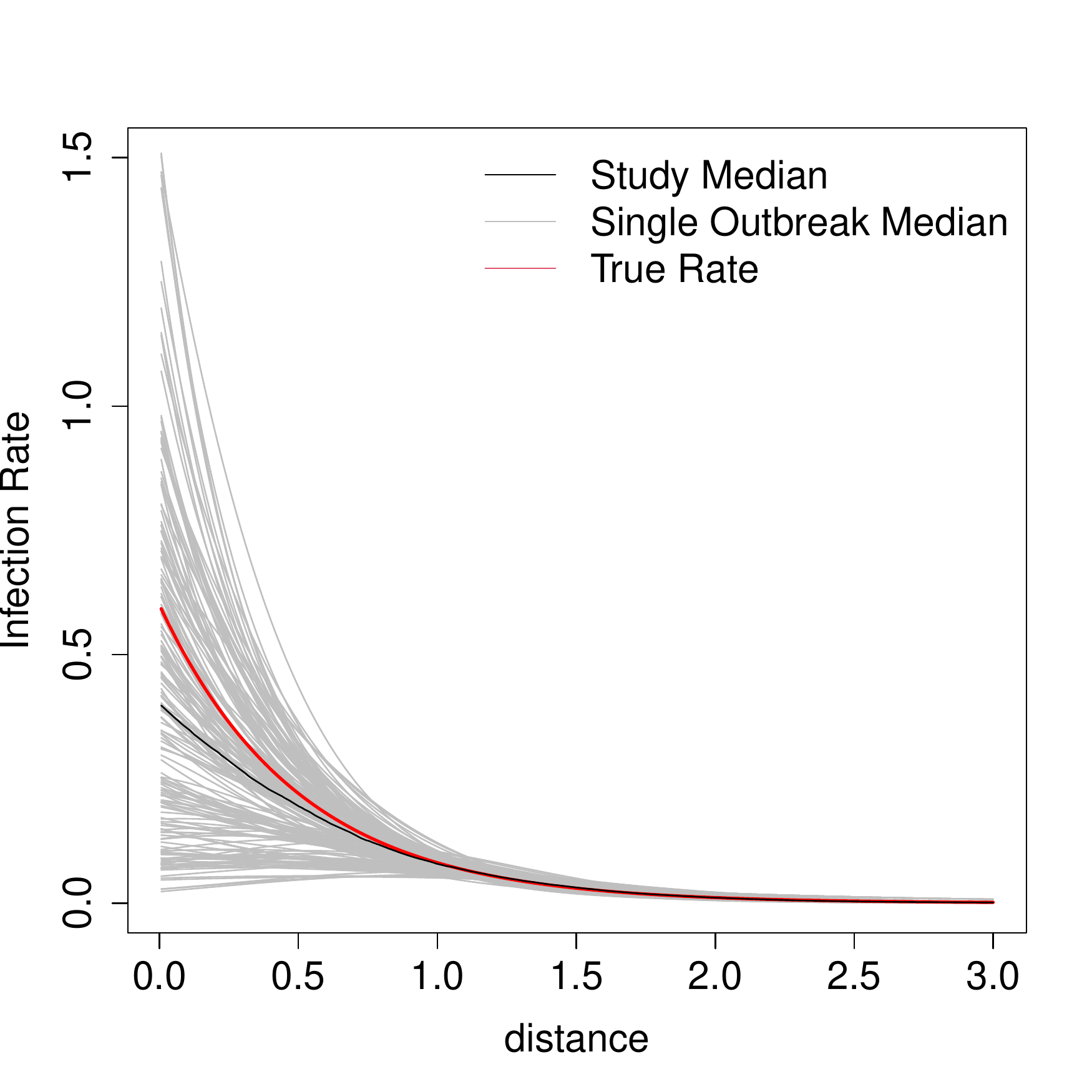}
\end{figure}

Figure \ref{fig: simulation study posterior rates} shows the median rate compared to the true rate and a 95\% credible interval constructed from all 175 posterior medians. The results demonstrate that we  can infer the infection rate function well for all pairwise distances above 0.5km, but we slightly underestimate the rate between immediate neighbours. This underestimate is caused by there being few farms in each data set that are less than 0.5km apart. We estimate the median infectious period to be 5.07 days, close to the true value of 5 days, and the 95\% credible interval of the 175 estimates contains the true value of 5. This slight overestimation is likely to be caused by the combination of slight underestimation of $\beta$ at low distances, and the fact that we only considered data sets with at least 100 infections, in which infectious periods may be slightly larger than average. 

To assess the results for the infection times across all simulations, we use the relative percentage error in the sum of the infection times for each simulated outbreak, which we denote by $\tilde{i}$, defined as
follows. Consider a single simulated data set. 
Let $S$ denote the sum of all infection times of farms culled either naturally or pre-emptively in the data set. Let $\hat{S}$ denote the median estimate of $S$ obtained from the MCMC output. Note that $\hat{S}$ implicitly takes account of which pre-emptively culled farms are imputed to have been infected. Then we define

$$
\tilde{i} = \frac{S - \hat{S}}{S} \times 100 \% .
$$
The median relative percentage error across all data sets is 1.40\%, which demonstrates that our method for inferring infection times gives accurate results. The results for the parameters are shown in Table \ref{tab: sim study results}.

\begin{table}
	\caption{\label{tab: sim study results} Summary statistics for the
	175 posterior median values obtained in the simulation study. The
	probability interval is from the 2.5\% to 97.5\% quantiles.}	
	\centering
	\begin{tabular}{|c|c|c|c|}
	\hline
	Parameter & True Value & Median & 95\% Prob. Int.  \\
	\hline
	$\gamma$ & 0.8 & 0.787 & (0.778, 0.802) \\
	$\lambda/ \gamma$ & 5 & 5.07 & (4.99, 5.13) \\
	$\tilde{i}$ & 0\% & 1.40\% & (-9.18\%, 23.0\%) \\ \hline 
	\end{tabular}
\end{table}

\subsection{Avian Influenza}
We now analyse the Avian Influenza data described in Section \ref{subsec:data}. Due to the size of the data set, we split the inference into two parts. We first inferred plausible values of the GP prior distribution length scale parameter, $l$, by fitting our transmission model under the assumptions of a constant infectious period of seven days and that pre-emptively culled farms were not infected, as in \citet{Elb07}. We obtained a posterior median for $l$ of 2.75km (95\% CI: (2.55, 3.01)). The reason for inferring plausible values for $l$ separately is that estimating $l$ requires decomposing and inverting the covariance matrix inside the MCMC algorithm which is highly computationally intensive and leads to prohibitively long run times. This issue is amplified when the infection times are unknown as well.

We repeated the inference method without assuming that the infection times or the status of the pre-emptively culled farms are known. Based on the results of the method with a fixed infectious period, we fixed $\alpha =3$ and $l = 3$km. We employed the GP approximation method for this data set. As we expect the infection rate function to vary considerably over short to medium distances, we included more such distances in the pseudo data set. The pseudo data set was $\boldsymbol{\bar{d}} =\{0, 0.5, 1, \ldots, 19.5, 20, 30, \ldots, 350\}$. We ran the MCMC algorithm for 20,000 iterations, including a burn-in period of 500 iterations. In each iteration of the MCMC algorithm, we proposed updating, adding or deleting 200 infection times. This took 7 days on the University of Nottingham's High Performance Computing facility.

\begin{figure}
	\caption{The posterior mean for the nonparametric (solid) and parametric (dashed) infection rate functions for the Avian Influenza data set. The parametric function is kernel 3 in Table \ref{tab: ai parametric rates}.}
	\label{fig: avian influenza posterior rates}
	\includegraphics[width = \textwidth]{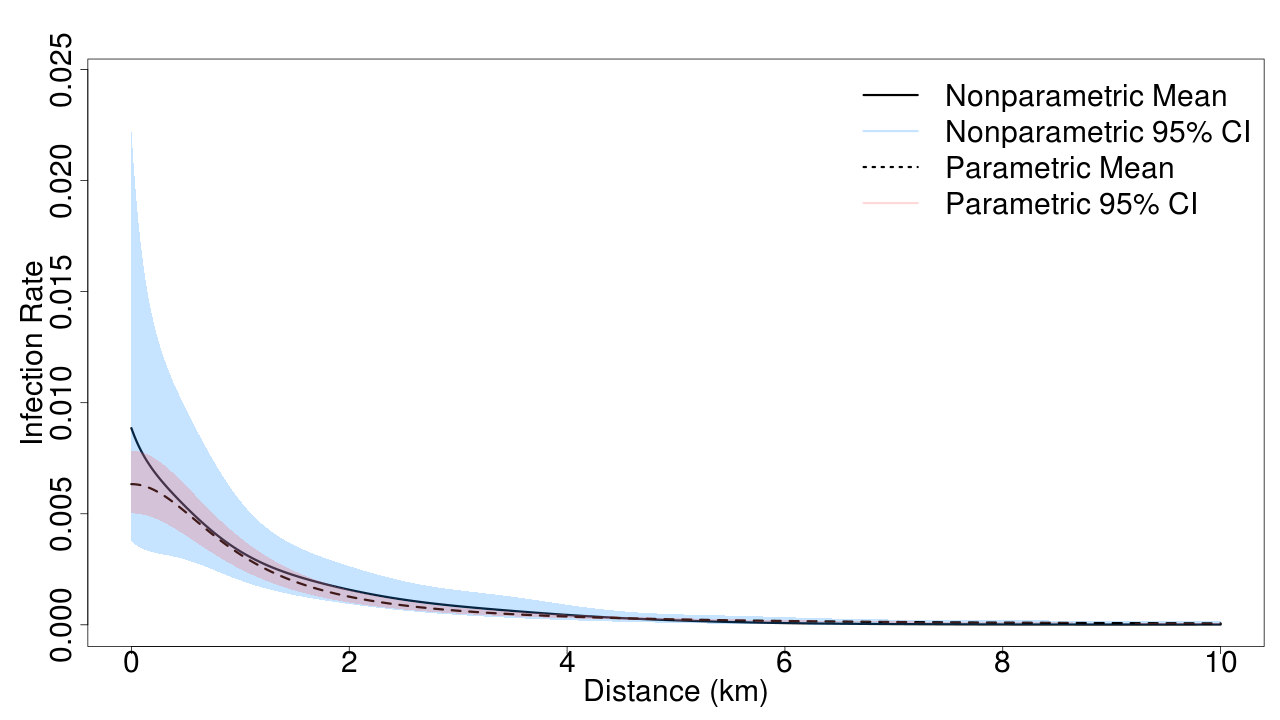}
\end{figure}

The results for the infection rate are shown in Figure \ref{fig: avian influenza posterior rates}, where we see a logistic-type function that decays to zero. From this, we estimate that the probability of a farm infecting another farm which is more than 6km apart is negligible. From the credible interval, we see that samples from the posterior distribution take a variety of shapes, with functions that have a high infection rate over short distance decaying quickly, and functions that have a lower rate over short distances taking a logistic function form. 

We compare our results to those in \cite{Elb07}, particularly with a view to comparing estimation of the infection rate function. The authors propose five models, shown in Table \ref{tab: ai parametric rates}, which we fitted to the data assuming a fixed infectious period of seven days. Model 3 was the best of the proposed models according to the Deviance Information Criterion \citep{DIC_paper}. We refitted model 3 to the data assuming that the infection times are unknown. The results are shown in Figure \ref{fig: avian influenza posterior rates} and one clear difference between the parametric and nonparametric methods is the associated uncertainty. Although the nonparametric method allows for a greater degree of flexibility, it also induces a greater degree of uncertainty. However, we argue that the parametric method may underestimate the uncertainty by imposing stricter assumptions. Despite this, both estimates are of similar shape and scale, and our results broadly agree with existing work. We see a slight difference in the forms of the infection rate function for distances less than 400m, which is due to there being very few farms that are less than 400m apart. 

\begin{figure}
\centering
	\caption{Top: Posterior distribution of median infectious periods for farms with confirmed infections and those pre-emptively culled. Bottom: Estimates of probabilities that pre-emptively culled farms were infected. Only farms which were pre-emptively culled are plotted. Each probability is the proportion of iterations in the MCMC algorithm that the pre-emptively culled farm was actually infected.}
	\label{fig: avian flu IPD}
	\includegraphics[width = 0.8\textwidth]{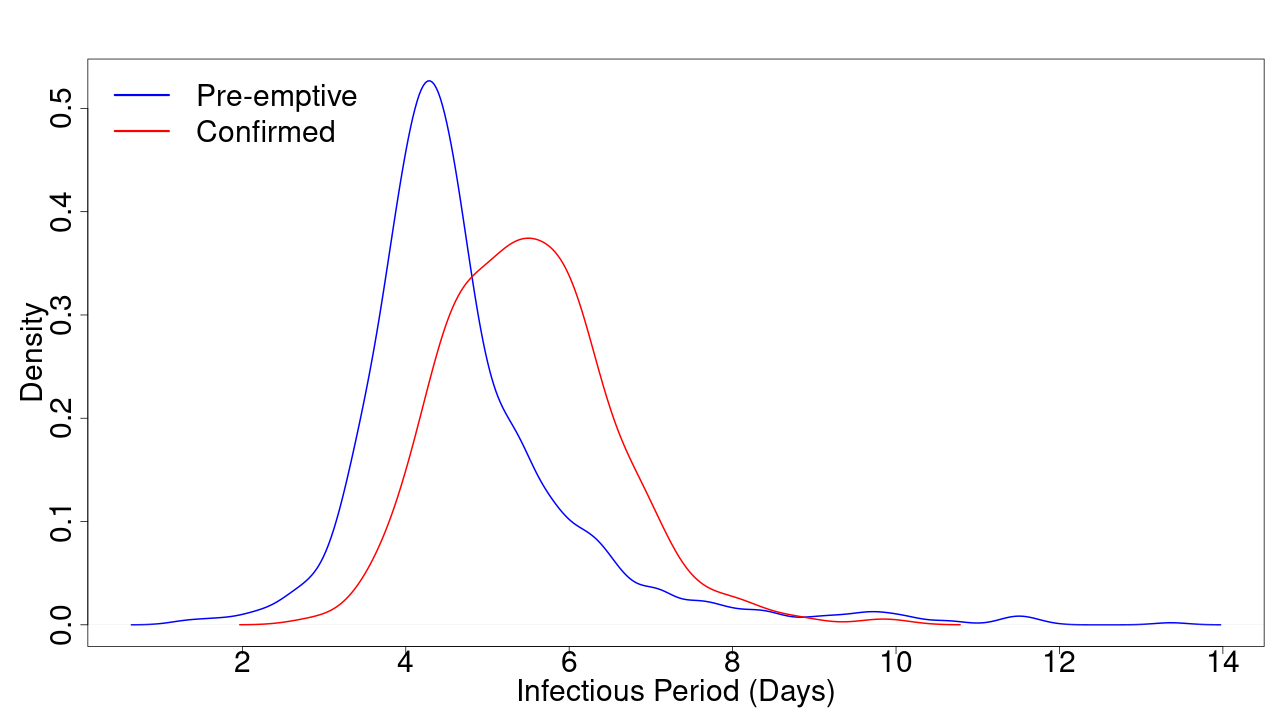} 
	\includegraphics[width =0.8\textwidth]{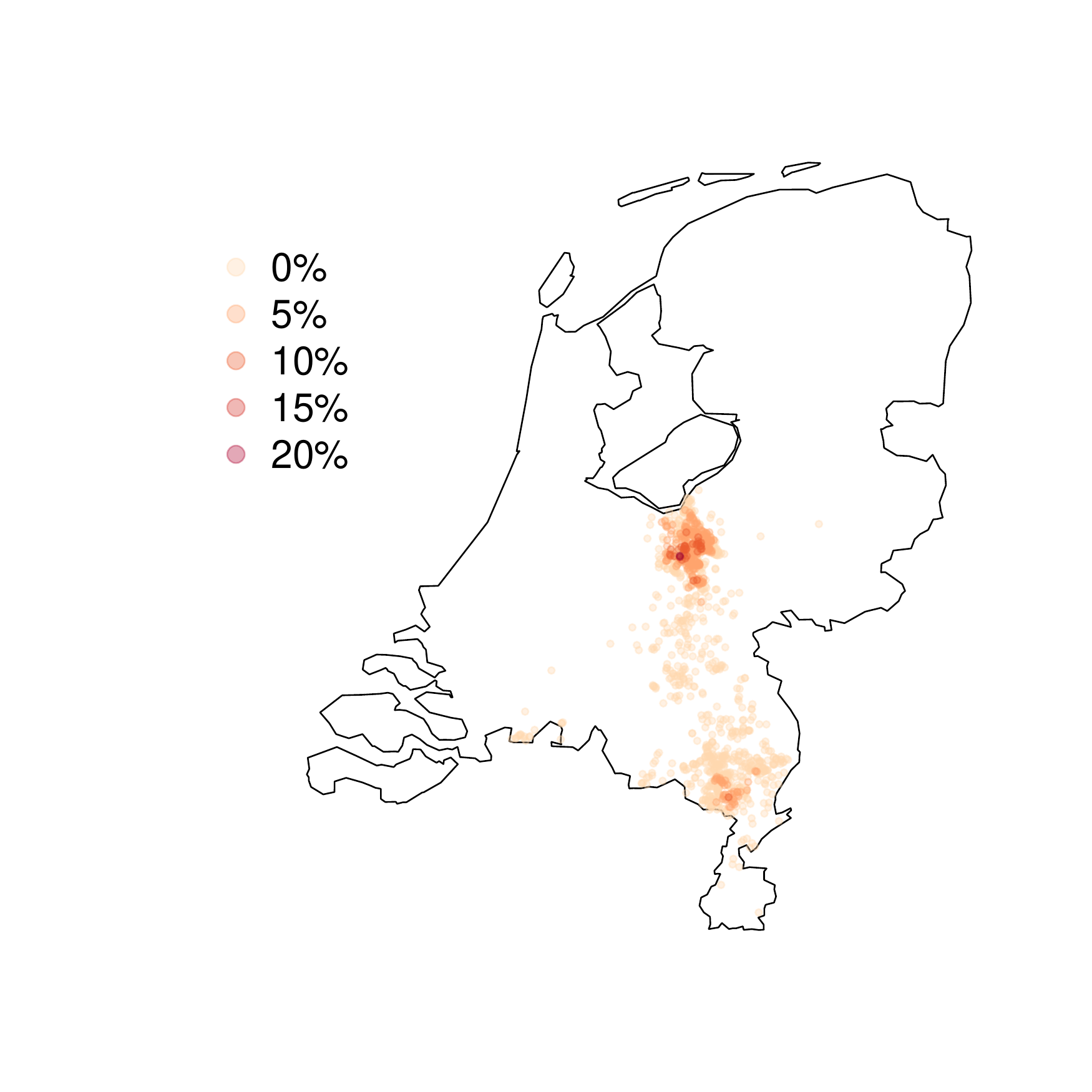}
\end{figure}

Since we assume infection times to be unknown, we infer them via our MCMC algorithm. We estimate the mean infectious period to be 6.4 days, and Figure \ref{fig: avian flu IPD} shows the distribution of median infectious periods by culling status. For farms that were subject to pre-emptive culling, the median infectious period is shorter than for those who were identified as infected. This is expected as pre-emptive culling of infected farms introduces censoring. 

We estimate the probability that each pre-emptively culled farm was actually infected, as shown in Figure \ref{fig: avian flu IPD}. All of the farms with non-zero probability of infection are located in the two main infection clusters. Our results show that the transmission to the southern cluster cannot be explained by a path of shorter distance infections that were censored by pre-emptive culling. This is consistent with the hypothesis proposed in \cite{Bataille2011} that this long distance transmission event of Avian Influenza was the result of a single human-mediated transport of the virus. 

\begin{table}
	\caption{\label{tab: ai parametric rates} The proposed parametric pair-wise infection rates for the Avian Influenza data set in \cite{Elb07}.}	
	\centering
	\begin{tabular}{|c|c|}
	\hline
	Rate & Kernel \\
	\hline
	1 & $\beta(d) = \beta_0$ \\
	2 & $\beta(d) = \beta_0(1+d)^{-1}$ \\
	3 & $\beta(d) = \beta_0(1+d^2)^{-1}$ \\
	4 & $\beta(d) = \beta_0(1+d^{\beta_1})^{-1}$ \\
	5 & $\beta(d) = \beta_0(1+(d/\beta_2)^{\beta_1})^{-1}$ \\\hline 
	\end{tabular}
\end{table}

\subsection{Culling Strategies}
We now investigate how to improve the disease control measures by analysing how the culling radius affects the number of infected farms. Culling infected farms has the effect of reducing the time a farm is infectious, and culling susceptible farms means there are fewer farms to be infected. Although this is an effective measure for controlling the spread of the disease, it can be expensive as farmers are compensated for lost livestock and it can cause negative public attitudes. 

To simulate the effect of culling, we sample from the posterior predictive distribution of the infection and culling times. Given the observed culling times, and the posterior distributions of $g$, $\gamma$ and $\omega$, we wish to generate new infection times $\textbf{i}^*$ for all farms, and corresponding culling times $\textbf{r}^*$. We do this using the posterior predictive distribution, which is given by
$$
\pi(\textbf{i}^*, \textbf{r}^*\mid \textbf{r}) = \iiint \pi(\textbf{i}^*, \textbf{r}^*\mid g, \gamma, \omega, \textbf{r})\pi(g, \gamma, \omega\mid \textbf{r})\hbox{d}g\,\hbox{d}\gamma\,d\omega.
$$
To generate samples from this distribution we initiate the outbreak by assuming the initially infected farm $\omega$ is the farm that was initially culled in the observed outbreak. To consider the effectiveness of culling strategies, we assume that once an infected farm reaches the end of its infectious period and enters the removed class all farms up to $r$km away are simultaneously culled and enter the removed class. Culling cannot start immediately as it may take time for the authorities to be notified of the disease and put measures into place, and whereas previous work \citep{Back15} uses a fixed delay after the first detection to initiate the culling measures, we allow for stochasticity in the disease take-off and assume culling takes place once a certain number of farms have been infected. As resources may not be immediately available to the authorities, it may not be possible to cull all farms within $r$km and we simulate this by fixing a maximum number of farms that can be culled per day. We then increase this number over the course of the outbreak as the authorities have more available resources. The numbers are given in Table \ref{tab: Culling Simulation Take Off} and are based on the number of farms we estimate to have been infected in the observed outbreak. Similarly, we assume the authorities will not have sufficient resources to cull all farms within the chosen radius at the start of the outbreak, and we model this by assuming they initially cull farms within a radius half as large. 
\begin{table}
\caption{\label{tab: Culling Simulation Take Off} The culling strategy as a function of the total number of infected farms.}
\centering
	\begin{tabular}{|c|c|c|} \hline
	\vtop{\hbox{\strut Total Number of}\hbox{\strut Infected Farms (I)}} & \vtop{\hbox{\strut Maximum Number of}\hbox{\strut Farms Culled per day}} & \vtop{\hbox{\strut Proportion of Culling}\hbox{\strut Radius Implemented}} \\ \hline
	 $I \le 33$ & 0 & 0\\
	 $ 33 \le I \leq 54$ & 3 & $\frac{1}{2}$\\
	 $54 < I$ & 6 & 1 \\ \hline
	\end{tabular}
\end{table}

To investigate the economic consequences of these strategies, we assume each farmer is compensated for their culled livestock. We use additional data from the outbreak which describes the type of poultry on each farm (broiler, duck, turkey and layer) and the number of birds on each farm. The value of the compensation depends on the type of bird culled, the number of birds culled, their age in weeks, and, for turkeys, their gender. We follow \cite{Back15} who use the approximate rates shown in Table \ref{tab: compensation Backer}. We acknowledge this method is crude and does not take into account any of the wider economic impacts. However, it allows us to simulate the number of farms that are infected, the number of farms that are culled, and the compensation paid to farmers. These three values can be used to compare the risk to public health, the impact of the poultry industry, and the cost to the authorities. 

\begin{table}
	\caption{\label{tab: compensation Backer} Estimates of compensation per bird paid to farmers during the Avian Influenza outbreak from \cite{Back15}.} 
	\centering
	\begin{tabular}{|c|c|} \hline
	Poultry Type & Compensation (\euro per bird) \\ \hline
	Broiler & 0.98 \\
	Duck & 2.09 \\
	Turkey & 10.63 \\
	Layer  & 2.05 \\
	\hline
	\end{tabular}
\end{table}

Table \ref{tab: posterior predictive results} shows the results of the culling strategies for radii between 0km and 5km. A culling radius of 0km denotes the authorities taking no action. It is clear that taking any course of action leads to a reduction of the number of infected farms but also an increase of the amount of compensation given. Furthermore, we see that more ambitious strategies show little gain in reducing the median number of farms infected in an outbreak. The effect of culling at larger radii results in a larger number of culled farms and a higher amount of compensation, but does not result in a considerable reduction in the number of infected farms. This is because the maximum number of farms culled per day is quickly reached, even for small culling radii. In the data set, the average density of farms was approximately 2 per km$^2$, whereas a culling radius of 2km covers over 12km$^2$.

\begin{table}
	\caption{\label{tab: posterior predictive results} Posterior predictive medians (95\% probability intervals) for the number of infected and culled farms and the amount of compensation paid. }
	\begin{tabular}{|c|c|c|c|} \hline
	Radius (km) & No. Infected Farms & No. Culled Farms & Compensation Paid (\euro millions) \\ \hline
	0 & 443 (151, 644) & 443 (151, 644) & 24.8 (8.62, 35.9)\\
	1 & 297 (110, 535) & 489 (215, 709) & 27.2 (12.2, 38.9)\\
	2 & 283 (108, 608) & 488 (217, 740) & 27.5 (12.2, 41.7)\\
	3 & 283 (112, 582) & 517 (242, 775) & 29.0 (13.2, 43.1)\\
	4 & 274 (105, 564) & 512 (228, 793) & 28.5 (12.3, 43.9)\\
	5 & 280 (109, 549) & 527 (226, 797) & 39.2 (12.4, 41.9)\\\hline
	\end{tabular}
\end{table}

These results are broadly in line with those in \cite{Back15}, which also suggest that larger culling radii do not result in a considerable reduction in the number of infected farms. However, as we use a much smaller estimate for the maximum number of farms culled per day, we do not observe a large difference between culling radii of 1km and 2km.

\section{Conclusions} \label{sec:conclusion}
We have presented an analysis of an outbreak of Avian Influenza in poultry farms in the Netherlands using a Bayesian nonparametric approach. Our approach demonstrates that it is possible to model the spatially heterogeneous infection rate for infectious diseases nonparametrically, and that GPs provide a flexible framework for doing so. This nonparametric methodology allows us to reduce the need for strict parametric assumptions, which are often made for mathematical or practical convenience and may have little scientific basis. Our methods also allow us to account for missing data, specifically the unobserved process of infection, without making unrealistic simplifying assumptions.

Although we have focused on an SIR model, in principle our methods can be extended to SEIR models as well, to incorporate a latent period. For our application to Avian Influenza, transmission experiments suggest that for the A/H7N7 virus in chickens, the latent period for an infected animal is between 1 and 2 days \citep{van2005}. The latent period of an infected farm is often equated to the latent period of the first infected chicken, i.e. in this case that would suggest a fairly short latent period of between 1 and 2 days \citep[e.g.][]{Back15}. Furthermore, in \citet{Ypma11} the authors assumed a latent period of 1 day and subsequently performed a sensitivity analysis,  where they compared results across latent periods of lengths 1, 2, and 3 days. Their comparison shows that their estimated kernel parameters were essentially insensitive to the assumed latent period duration. Although we could have considered fitting an SEIR model with a short latent period, we anticipate that this would not have any material impact on our results.

The methods we have described require more time and computational power than the standard parametric methods, especially when employed in conjunction with an MCMC approach to sample from the desired posterior distribution. We have however somewhat alleviated these issues by using a GP approximation method which appears to work well in our applications. Simulation studies in \citet{Seymour20} suggest that our methods work well even in small populations (e.g. $N=100$), although there needs to be enough transmission in the population leading to a sizeable outbreak. Conversely, in scenarios where fewer data are available, such as small outbreaks or the initial phases of an outbreak, then in common with any nonparametric approach there will be greater uncertainty in parameter estimates.  In such situations it might be appropriate to incorporate strong prior information or simply revert to a parametric approach.

For the Avian Influenza data set, our methodology has allowed us to approach the infection process in a more flexible way than previous methods. Our estimates are in line with previous work, and combining this method with previously developed MCMC techniques and data augmentation allows us to analyse this data set in more detail than has previously been possible, including determining whether pre-emptively culled farms had been infected. The uncertainty around our estimates is larger than that of previous parametric methods, but since we do not assume specific parametric models then our methods are, in some sense, giving a fairer quantification of uncertainty. We are able to use the posterior predictive distribution to analyse the effect of different control strategies which can be used to inform policy in this area. 

In this paper, we have focussed on spatial heterogeneity as the key determinant of the infection rate. In reality, it is possible that the number and type of animals on the farms was also important. Given appropriate data, it is natural to build a model which contains such data as covariates. One way of doing this would be to consider each covariate as a separate dimension of the GP. Another possible extension is to consider different covariance functions beyond the squared exponential function, which could be appropriate in some applications. Also, the proposed framework can be extended to analyse the spread of infectious diseases early on in an outbreak. As mentioned above, the lack of data in the initial stages of an outbreak may be problematic. Possible ways to mitigate this by include adding further assumptions to the model, such as monotonicity of the infection rate, as well as employing more informative prior distributions. It is also be of interest to develop methods for model assessment, which is something that we have not considered here. Finally, another avenue for future work is to employ the recently developed methods described in \citet{StockKypOn19} in which the observed data likelihood is approximated without the need for imputing the unknown infection times. This would significantly reduce the computation time needed for our methods, and in conjunction with the GP projection method can make the proposed methodology more applicable for very large data sets. 
  
\section{Acknowledgements}
We thank Wageningen Bioveterinary Research, The Netherlands Food and Consumer Product Authority and the Dutch Ministry of Agriculture, Nature and Food Quality for sharing anonymised outbreak, culling and denominator data of the Dutch 2003 HPAI epidemic with us. 

We are grateful for access to the University of Nottingham High Performance Computing Facility. 

We also thank the Associate Editor and the two reviewers for helpful and constructive comments that have improved this article.

This work was supported by the UK Engineering and Physical Sciences Research Council (EPSRC) grant EP/N50970X/1.

\bibliographystyle{apalike}
\bibliography{bibliography}

\begin{thebibliography}{}

\bibitem[Andersson and Britton, 2000]{And00}
Andersson, H. and Britton, T. (2000).
\newblock {\em Stochastic epidemic models and their statistical analysis}.
\newblock Lecture Notes in Statistics. Springer.

\bibitem[Backer et~al., 2015]{Back15}
Backer, J., van Roermund, H., Fischer, E., van Asseldonk, M., and Bergevoet, R.
  (2015).
\newblock Controlling highly pathogenic avian influenza outbreaks: An
  epidemiological and economic model analysis.
\newblock {\em Preventive Veterinary Medicine}, 121(1-2):142--150.

\bibitem[Bailey, 1975]{Bail75}
Bailey, N. T.~J. (1975).
\newblock {\em The mathematical theory of infectious diseases and its
  applications}.
\newblock Griffin, London, 2nd edition.

\bibitem[Bataille et~al., 2011]{Bataille2011}
Bataille, A., van~der Meer, F., Stegeman, A., and Koch, G. (2011).
\newblock Evolutionary analysis of inter-farm transmission dynamics in a highly
  pathogenic avian influenza epidemic.
\newblock {\em {PLoS} Pathogens}, 7(6):e1002094.

\bibitem[Bavinck et~al., 2009]{Bavinck2009}
Bavinck, V., Bouma, A., van Boven, M., Bos, M., Stassen, E., and Stegeman, J.
  (2009).
\newblock The role of backyard poultry flocks in the epidemic of highly
  pathogenic avian influenza virus ({H}7{N}7) in the netherlands in 2003.
\newblock {\em Preventive Veterinary Medicine}, 88(4):247--254.

\bibitem[Becker, 1989]{Beck89b}
Becker, N.~G. (1989).
\newblock {\em Analysis of infectious disease data}.
\newblock Chapman and Hall, London.

\bibitem[Boender et~al., 2007]{Elb07}
Boender, G.~J., Hagenaars, T.~J., Bouma, A., Nodelijk, G., Elbers, A. R.~W.,
  de~Jong, M. C.~M., and van Boven, M. (2007).
\newblock Risk maps for the spread of highly pathogenic avian influenza in
  poultry.
\newblock {\em {PLoS} Computational Biology}, 3(4):e71.

\bibitem[Chalupka et~al., 2013]{chalupka2013}
Chalupka, K., Williams, C.~K., and Murray, I. (2013).
\newblock A framework for evaluating approximation methods for gaussian process
  regression.
\newblock {\em Journal of Machine Learning Research}, 14:333--350.

\bibitem[Csato and Opper, 2002]{Csat02}
Csato, L. and Opper, M. (2002).
\newblock Sparse online {G}aussian processes.
\newblock {\em Neural Computation}, 14(3):641--668.

\bibitem[{Directorate-General for Health and Consumers}, 2003]{DGSanco03}
{Directorate-General for Health and Consumers} (2003).
\newblock Avian influenza ({AI}) in the {N}etherlands, {B}elgium and {G}ermany
  – chronology of main events and list of decisions adopted by the
  commission.
\newblock Technical report, European Commission.

\bibitem[Elbers et~al., 2004]{Elb04}
Elbers, A. R.~W., Fabri, T. H.~F., de~Vries, T.~S., de~Wit, J.~J., Pijpers, A.,
  and Koch, G. (2004).
\newblock The highly pathogenic avian influenza {A} ({H7N7}) virus epidemic in
  the netherlands in 2003{\textemdash}lessons learned from the first five
  outbreaks.
\newblock {\em Avian Diseases}, 48(3):691--705.

\bibitem[Fouchier et~al., 2004]{Fouch04}
Fouchier, R. A.~M., Schneeberger, P.~M., Rozendaal, F.~W., Broekman, J.~M.,
  Kemink, S. A.~G., Munster, V., Kuiken, T., Rimmelzwaan, G.~F., Schutten, M.,
  van Doornum, G. J.~J., Koch, G., Bosman, A., Koopmans, M., and Osterhaus, A.
  D. M.~E. (2004).
\newblock Avian influenza {A} virus ({H7N7}) associated with human
  conjunctivitis and a fatal case of acute respiratory distress syndrome.
\newblock {\em Proceedings of the National Academy of Sciences},
  101(5):1356--1361.

\bibitem[Hensman et~al., 2013]{Hens13}
Hensman, J., Fusi, N., and Lawrence, N.~D. (2013).
\newblock {G}aussian processes for big data.
\newblock In {\em Conference on Uncertainty in Artificial Intelligence}, pages
  282--290.

\bibitem[Jewell et~al., 2009]{Jew09}
Jewell, C.~P., Kypraios, T., Neal, P., and Roberts, G.~O. (2009).
\newblock Bayesian analysis for emerging infectious diseases.
\newblock {\em Bayesian Analysis}, 4(3):465--496.

\bibitem[Koopmans et~al., 2004]{Koop04}
Koopmans, M., Wilbrink, B., Conyn, M., Natrop, G., van~der Nat, H., Vennema,
  H., Meijer, A., van Steenbergen, J., Fouchier, R., Osterhaus, A., and Bosman,
  A. (2004).
\newblock Transmission of {H7N7} avian influenza {A} virus to human beings
  during a large outbreak in commercial poultry farms in the netherlands.
\newblock {\em The Lancet}, 363(9409):587--593.

\bibitem[Neal, 1998]{Neal98}
Neal, R. (1998).
\newblock Regression and classification using {G}aussian process priors.
\newblock In Bernardo, J.~M., Berger, J.~O., Dawid, A.~P., and Smith, A. F.~M.,
  editors, {\em Bayesian Statistics 6}. Oxford Univeristy Press.

\bibitem[O'Neill and Roberts, 1999]{On99}
O'Neill, P.~D. and Roberts, G.~O. (1999).
\newblock Bayesian inference for partially observed stochastic epidemics.
\newblock {\em Journal of the Royal Statistical Society: Series A},
  162(1):121--129.

\bibitem[Quinonero-Candela and Rasmussen, 2005]{Quin05}
Quinonero-Candela, J. and Rasmussen, C.~E. (2005).
\newblock A unifying view of sparse approximate {G}aussian process regression.
\newblock {\em Journal of Machine Learning Research}, 6.

\bibitem[Rasmussen and Williams, 2006]{Ras06}
Rasmussen, C.~E. and Williams, C. (2006).
\newblock {\em {G}aussian Processes for Machine Learning}.
\newblock MIT Press.

\bibitem[Seymour, 2020]{Seymour20}
Seymour, R.~G. (2020).
\newblock {\em Bayesian nonparametric methods for individual-level stochastic
  epidemic models}.
\newblock PhD thesis, University of Nottingham.

\bibitem[Spiegelhalter et~al., 2002]{DIC_paper}
Spiegelhalter, D.~J., Best, N.~G., Carlin, B.~P., and Van Der~Linde, A. (2002).
\newblock Bayesian measures of model complexity and fit.
\newblock {\em Journal of the Royal Statistical Society: Series {B}},
  64(4):583--639.

\bibitem[Stegeman et~al., 2004]{Bov03}
Stegeman, A., Bouma, A., Elbers, A.~R.~W., de~Jong, M.~C.~M., Nodelijk, G.,
  de~Klerk, F., Koch, G., and van Boven, M. (2004).
\newblock Avian influenza {A} virus ({H7N7}) epidemic in the netherlands in
  2003: Course of the epidemic and effectiveness of control measures.
\newblock {\em The Journal of Infectious Diseases}, 190(12):2088--2095.

\bibitem[Stockdale et~al., ress]{StockKypOn19}
Stockdale, J.~E., Kypraios, T., and O’Neill, P.~D. (in press).
\newblock Pair-based likelihood approximations for stochastic epidemic models.
\newblock {\em Biostatistics}, page
  {https://doi.org/10.1093/biostatistics/kxz053}.

\bibitem[Van~der Goot et~al., 2005]{van2005}
Van~der Goot, J., Koch, G.~d., De~Jong, M., and Van~Boven, M. (2005).
\newblock Quantification of the effect of vaccination on transmission of avian
  influenza ({H}{7}{N}7) in chickens.
\newblock {\em Proceedings of the National Academy of Sciences},
  102(50):18141--18146.

\bibitem[van~der Goot et~al., 2005]{Goot05}
van~der Goot, J.~A., Koch, G., de~Jong, M. C.~M., and van Boven, M. (2005).
\newblock Quantification of the effect of vaccination on transmission of avian
  influenza ({H7N7}) in chickens.
\newblock {\em Proceedings of the National Academy of Sciences},
  102(50):18141--18146.

\bibitem[Ypma et~al., 2011]{Ypma11}
Ypma, R. J.~F., Bataille, A. M.~A., Stegeman, A., Koch, G., Walling, J., and
  van Ballegooijen, W.~M. (2011).
\newblock Unravelling transmission trees of infectious diseases by combining
  genetic and epidemiological data.
\newblock {\em Proceedings of the Royal Society {B}}, 279(1728):444--450.

\bibitem[Zhang, 2004]{Zhang2004}
Zhang, H. (2004).
\newblock Inconsistent estimation and asymptotically equal interpolations in
  model-based geostatistics.
\newblock {\em Journal of the American Statistical Association},
  99(465):250--261.

\end{thebibliography}

\end{document}


\maketitle

\section{Metropolis-Hastings Ratio When Updating the Infection Rate} 
We show that by using the proposal in Section 3.2.1 the Metropolis-Hastings ratio reduces to the likelihood ratio. The acceptance probability is given by
$$
p_{acc} = \frac{\pi(\textbf{i}, \textbf{r}^c, \mathcal{B}, \mathcal{C}, \mathcal{D} \mid g^\prime, \lambda, \gamma, \omega, i_\omega, \textbf{r}^p)}{\pi(\textbf{i}, \textbf{r}^c, \mathcal{B}, \mathcal{C}, \mathcal{D}\mid g, \lambda, \gamma, \omega, i_\omega, \textbf{r}^p)}\frac{\mathcal{GP}(g^\prime; 0, \Sigma)}{\mathcal{GP}(g; 0, \Sigma)}\frac{q(g|g^\prime)}{q(g^\prime|g)} \wedge 1,
$$
where $q(g^\prime|g)$ denotes the probability density of proposing $g^\prime(\textbf{d})$ given the current state $g(\textbf{d})$, both evaluated for a vector of distances $\textbf{d}$ in the domain of $g$. We also denote by $\mathcal{GP}(g; m, S)$ the probability density function of a multivariate normal distribution with mean vector $m$ and covariance matrix $S$ evaluated at $g(d)$, where $\textbf{d}$ is a vector of distances of in the domain of $g$.

\begin{eqnarray*}
	q\left(g^\prime|g\right) &=& \mathcal{GP}\left(g^\prime; \sqrt{1-\delta^2}g, \delta^2\Sigma \right) \\ 
	&=& \mathcal{GP}\left(\frac{1}{\delta}\left(\sqrt{1-\delta^2}\,g - g^\prime\right); 0, \Sigma\right) \\ 
	& \propto& \exp\left\{-\frac{1}{\delta^2} \left(\sqrt{1-\delta^2}\, g - g^\prime\right)^T\Sigma^{-1}\left(\sqrt{1-\delta^2}\, g - g^\prime\right)\right\} \\ 
	& \propto & \exp\left\{-\frac{1}{\delta^2} \left[\left(1-\delta^2\right)\,g^T\Sigma^{-1}g -  \sqrt{1-\delta^2}\left(g^T\Sigma^{-1}g^\prime\right) \right. \right. \\
	& - &  \left. \left. \sqrt{1-\delta^2}\,\left(g^{\prime^T}\Sigma^{-1}g\right) +  g^T\Sigma^{-1}g^\prime\right] \right\}.
\end{eqnarray*}

Similarly,
\begin{eqnarray*}
q\left(g|g^\prime\right) & \propto & \exp\left\{-\frac{1}{\delta^2} \left[(1-\delta^2)g^{\prime^T}\Sigma^{-1}g^\prime - \sqrt{1-\delta^2}\left(g^{\prime^T}\Sigma^{-1}g\right) \right. \right. \\
& - & \left. \left. \sqrt{1-\delta^2}\left(g^T\Sigma^{-1}g^\prime\right) +  g^{\prime^T}\Sigma^{-1}g\,\right]\right\}
\end{eqnarray*}

Hence, plugging in the expressions above for $q\left(g|g^\prime\right)$ and $q\left(g^\prime|g\right)$ in the proposal ratio, the latter reduces to the inverse of the prior density ratio:

\begin{eqnarray*}
	\frac{q\left(g|g^\prime\right)}{q(g^\prime|g)} & = & \frac{\exp\left\{-\frac{1}{\delta^2} \left[(1-\delta^2)\, g^{\prime^T}\Sigma^{-1}g^\prime + g^{\prime^T}\Sigma^{-1}g\,\right]\right\}}{\exp\left\{-\frac{1}{\delta^2} \left[(1-\delta^2)\,g^T\Sigma^{-1}g +  g^T\Sigma^{-1}g^\prime\right]\right\}} \\
	& = & \frac{\exp\left\{-g^T \Sigma^{-1}g\right\}}{\exp\left\{-{g^\prime}^T \Sigma^{-1}g^\prime\right\}} \\
	& = & \frac{\mathcal{GP}(g; 0, \Sigma)}{\mathcal{GP}(g^\prime; 0, \Sigma)}.
\end{eqnarray*} 

\section{Sensitivity Analysis on the Length Scale Parameter}
We have repeated the simulation study of Section 4.1 considering two additional values for the length scale parameter $l$, namely $l=2$ and $l=5$. That is, we assumed that the length scale parameter is known and fixed to $2$ and then fitted the same model to the 175 data sets to which we previously fitted the model assuming that $l=3$. We repeated the same procedure for $l=5$. 

\begin{figure}[H]
     \centering
     \includegraphics[width = 0.45\textwidth]{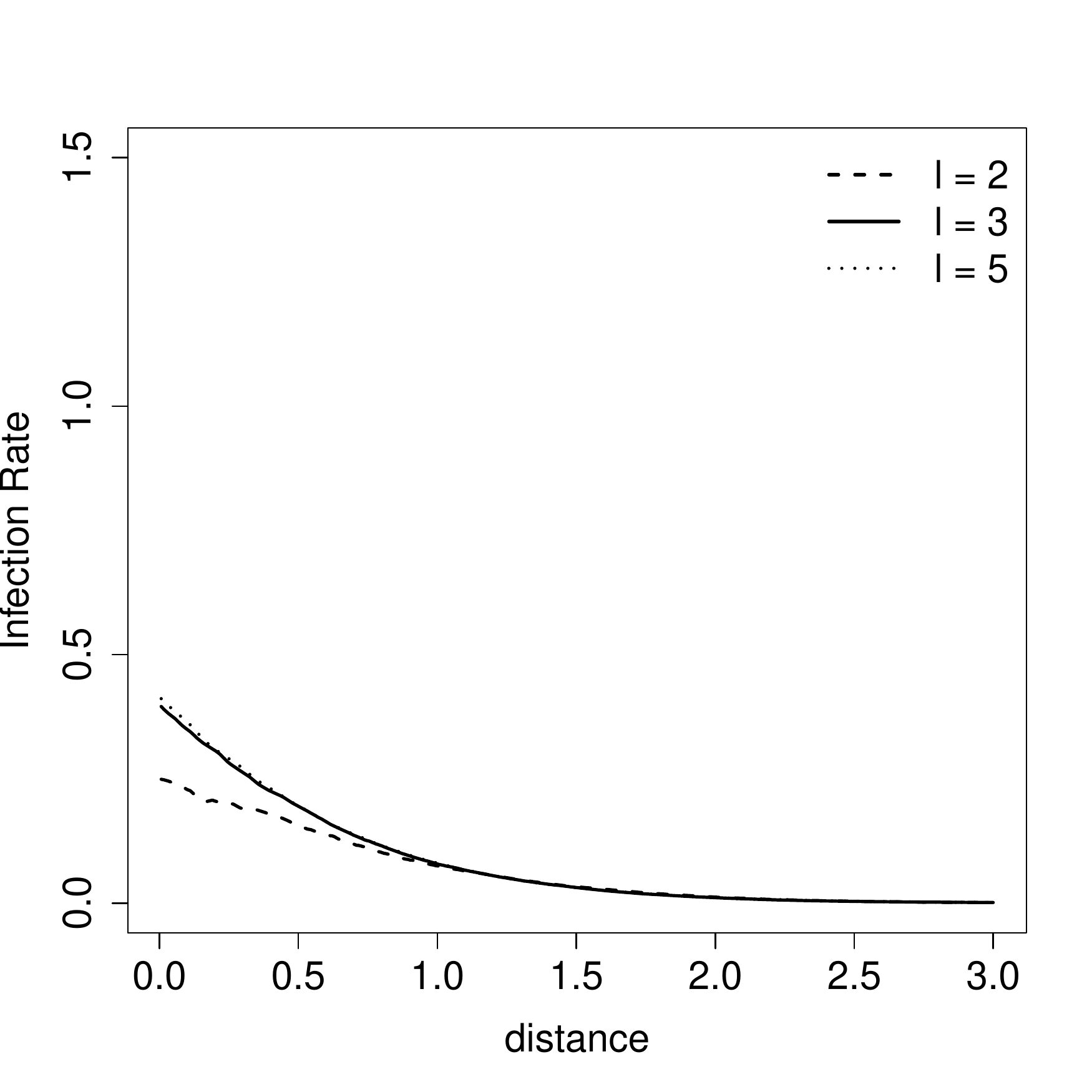}
     \includegraphics[width = 0.45\textwidth]{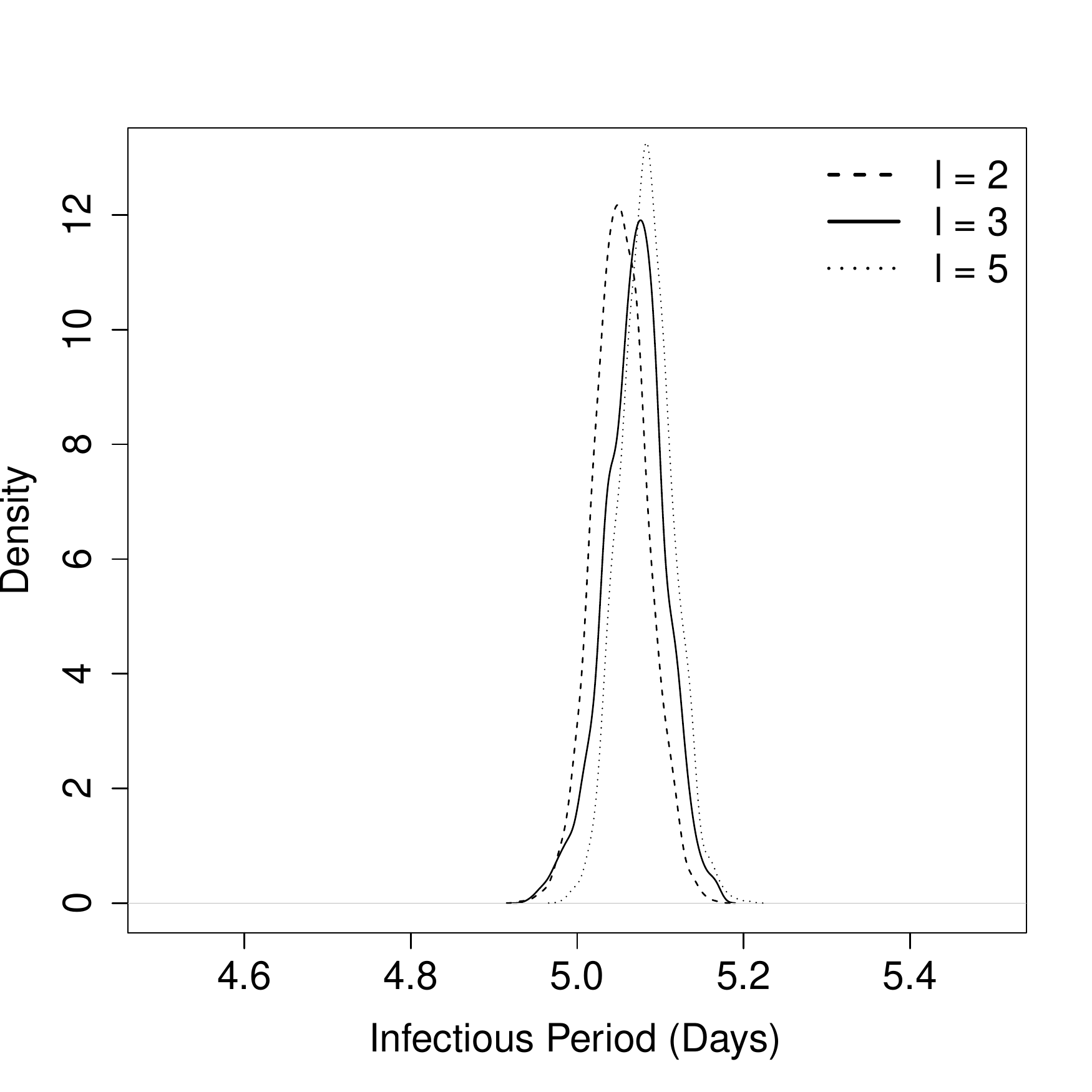}
     \caption{Left: Medians of the posterior median infection rates over 175 data sets. Right: The distribution of the posterior median estimates for the mean infectious period.} 
     \label{fig: sensitivity_lengthscale}
\end{figure}

Figure \ref{fig: sensitivity_lengthscale} shows that there is no material difference between the estimated infection rates for $l=3$ and $l=5$. Also, although there is some difference in the estimated curves between $l=2$ and $l=3$ for distances less than 0.5km, this is likely to be caused by there being few farms in each of the 175 data sets that are less than 0.5km apart. Furthermore, Figure \ref{fig: sensitivity_lengthscale} also shows that the inferred mean infectious period is fairly insensitive to the different choices of $l$.

